%% file: arXiv.tex
\documentclass[sigconf]{acmart}
\AtBeginDocument{%
  \providecommand\BibTeX{{%
    \normalfont B\kern-0.5em{\scshape i\kern-0.25em b}\kern-0.8em\TeX}}}

\setcopyright{acmlicensed}
\copyrightyear{2024}
\acmYear{2024}
\acmDOI{XXXXXXX.XXXXXXX}

\acmConference[Conference acronym 'XX]{Make sure to enter the correct
  conference title from your rights confirmation emai}{June 03--05,
  2024}{Woodstock, NY}
\acmISBN{978-1-4503-XXXX-X/18/06}
\input{lab-header}

\begin{document}

% \title{\AlgNameAbb: A Novel Method for Vulnerability Detection and Verification Rule Optimization}
\title{\AlgNameAbb: An Efficient SAT-Based Bayesian Optimization Algorithm for Verification Rules High-Risk Vulnerability Detection}

% \author{author: anonymous}
% \affiliation{
%  \institution{institution}
%  \country{country}
% }
% \email{email}

\author{Mao Luo}
\affiliation{%
  \institution{Hubei University of Technology\\
  Huazhong University of Science and Technology}
  \country{China}
}
\email{luomao@hbut.edu.cn}

\author{Zhi Wang}
\affiliation{%
  \institution{Ant Group}
  \country{China}
}
\email{wangchun.wz@antgroup.com}

\author{Yiwen Huang}
\affiliation{%
  \institution{Huazhong University of Science and Technology}
  \country{China}
}
\email{yiwen_huang@hust.edu.cn}

\author{Qingyun Zhang}
\affiliation{%
  \institution{Huazhong University of Science and Technology}
  \country{China}
}
\email{qingyun_zhang@hust.edu.cn}

\author{Zhouxing Su †}
\affiliation{%
  \institution{Huazhong University of Science and Technology}
  \country{China}
}
\email{suzhouxing@hust.edu.cn}

\author{Zhipeng Lü}
\affiliation{%
  \institution{Huazhong University of Science and Technology}
  \country{China}
}
\email{zhipeng.lv@hust.edu.cn}

\author{Wen Hu}
\affiliation{%
  \institution{Ant Group}
  \country{China}
}
\email{huwen.hu@antgroup.com}

\author{Jianguo Li †}\thanks{† Corresponding authors}
\affiliation{%
  \institution{Ant Group}
  \country{China}
}
\email{ljg263512@antfin.com}

\renewcommand{\shortauthors}{Trovato and Tobin, et al.}

\begin{abstract}
Electronic payment platforms are estimated to process billions of transactions daily, with the cumulative value of these transactions potentially reaching into the trillions. Even a minor error within this high-volume environment could precipitate substantial financial losses. To mitigate this risk, manually constructed verification rules, developed by domain experts, are typically employed to identify and scrutinize transactions in production environments. However, due to the absence of a systematic approach to ensure the robustness of these verification rules against vulnerabilities, they remain susceptible to exploitation. 

To mitigate this risk, manually constructed verification rules, developed by domain experts, are typically employed to identify and scrutinize transactions in production environments. However, due to the absence of a systematic approach to ensure the robustness of these verification rules against vulnerabilities, they remain susceptible to exploitation. To ensure data security, database maintainers usually compose complex verification rules to check whether a query/update request is valid. However, the rules written by experts are usually imperfect, and malicious requests may bypass these rules. As a result, the demand for identifying the defects of the rules systematically emerges. To address this challenging problem, we designed a SAT-Based Bayesian Optimization (SAT-BO) algorithm. SAT-BO encodes the problem into Boolean satisfiability problems (SAT) and searches for preferred truth value assignment by a customized DPLL algorithm. Based on the above SAT solver, we utilize a Bayesian optimization algorithm to determine the preference of variable assignment which corresponds to a large amount of requests. Experiment results show that our proposed algorithm can obtain solutions with high interception rates on several simulated scenarios, which verifies the effectiveness of SAT-BO on the generation of more comprehensive verification rules. In addition, we evaluate and justify the importance of three independent variable flipping strategies by experimental analysis.
% Consequently, malicious transactions may circumvent these expert-devised rules, resulting in significant financial detriment.
% \xadd{
% Vulnerability detection is an important topic for online database
% systems. To ensure data security, database maintainers usually compose complex verification rules to check whether a query/update request is valid. However, the rules written by experts are usually imperfect, and malicious requests may bypass these rules. As a result, the demand for identifying the defects of the rules systematically emerges. To address this challenging problem, we designed a SAT-Based Bayesian Optimization (SAT-BO) algorithm. SAT-BO encodes the problem into Boolean satisfiability problems (SAT) and searches for preferred truth value assignment by a customized DPLL algorithm. Based on the above SAT solver, we utilize a Bayesian optimization algorithm to determine the preference of variable assignment which corresponds to a large amount of requests. Experiment results show that our proposed algorithm can obtain solutions with high interception rates on several simulated scenarios, which verifies the effectiveness of SAT-BO on the generation of more comprehensive verification rules. In addition, we evaluate and justify the importance of three independent variable flipping strategies by experimental analysis.
% }
\end{abstract}





\maketitle
\setlength{\abovecaptionskip}{3pt}

\input{lab-body}

%\clearpage
\appendix
\section{Terminology}
Table \ref{tab:terms} presents the definitions of the key technical terms highlighted in bold in this paper.
\setlength{\abovecaptionskip}{3pt}
\setlength{\belowcaptionskip}{0pt}

\begin{table}[!h]
\centering
    \footnotesize
    %\scriptsize
    \begin{tabular}{lp{20em}}
    \toprule
     Term & Meaning \\
    \midrule
business principle & Summarized by experts reflecting business logic \\
verification rules & Logic expression derived from business principles to scrutinize transactions\\
vulnerability & Transaction violate business principle but not detected by verification rules\\
vulnerability detection & Discover the vulnerability before it causes financial loss\\
vulnerability attack& Generate synthetic transaction expecting to be vulnerability \\
business field& The field in the transaction\\
tamper field& The field for attack to tamper for generating synthetic transaction\\
single-field tampering approach& Only utilize single tamper field in a attack\\
attack rule& Logic expression generated from SAT-BO showing how to tamper the value in the tamper field \\
vulnerability coverage& Quantify the proportion of transactions that are affected
by the specified attack rule\\
DPLL& Stands for Davis-Putnam-Logemann-Loveland SAT complete algorithm framework\\
WalkSAT& A local search algorithm for solving SAT problems\\
CDCL& The conflict-driven clause learning SAT algorithm framework based on DPLL\\
  \bottomrule
    \end{tabular}
    \caption{Table of terms and corresponding meaning.}
    \label{tab:terms}
\end{table}

% \section{Relevance to Web Services}
% Electronic payment has become an essential functionality in many web services, such as Taobao and Pinduoduo. Consequently, enhancing the security capabilities of electronic transactions to safeguard users' financial assets has become critically important. Therefore, we can confidently assert that SAT-BO addresses a significant challenge in web services.
\setlength{\abovecaptionskip}{0pt}
\setlength{\belowcaptionskip}{0pt}
\section{Detail Experimental Data}
Here, we present detailed experimental data corresponding to section \ref{sec:offline} in the main text as an appendix, corresponding to \Tab{s} \ref{tab:Binomial} and \ref{tab:Power-Law}.
\setlength{\abovecaptionskip}{0pt}
\setlength{\belowcaptionskip}{0pt}
% These detailed data demonstrate the efficiency and robustness of the SAT-BO algorithm in solving the vulnerability detection and verification rule optimization problem.
\begin{table}[!h]
\centering
    \footnotesize
    %\scriptsize
    \begin{tabular}{lccccc}
    \toprule
     Instance &v62-c124&v68-c121&v69-c135&v75-c286&v91-c218 \\
     \midrule
     \#sol& 280&278&427&21&43\\ \midrule
     Instance & v94-c215&v100-c258&other 23 instances \\ \midrule
     \#sol& 41&16&less than 10& \\

  \bottomrule
    \end{tabular}
    \caption{The number of SAT solutions obtained by the Random solver within 150 seconds while the DPLL SAT solver can obtain at least 30 solutions in 1 second.}
    \label{tab:Random}
\end{table}
\setlength{\abovecaptionskip}{0pt}
\setlength{\belowcaptionskip}{0pt}

\begin{table*}[!tb]
    \centering
        %\setlength{\tabcolsep}{0.3pt}
        %\renewcommand{\arraystretch}{0.95}
        %\footnotesize
        \scriptsize
	\begin{tabular}{lrrrrrrrrrrrrrr}
		\toprule
		\multirow{2}*{Instance}&\multicolumn{2}{c}{\AlgNameAbb{0}}&\multicolumn{2}{c}{\AlgNameAbb{1}}&\multicolumn{2}{c}{\AlgNameAbb{2}}&\multicolumn{2}{c}{\AlgNameAbb{3}}&\multicolumn{2}{c}{\AlgNameAbb{4}}&\multicolumn{2}{c}{WalkSAT-BO}&\multicolumn{2}{c}{\AlgNameAbb{}}\\
		\cmidrule(lr){2-3}		\cmidrule(lr){4-5}		\cmidrule(lr){6-7}		\cmidrule(lr){8-9}		\cmidrule(lr){10-11}		\cmidrule(lr){12-13}    \cmidrule(lr){14-15}
		~&$f_{best}$&$f_{avg}$&$f_{best}$&$f_{avg}$&$f_{best}$&$f_{avg}$&$f_{best}$&$f_{avg}$&$f_{best}$&$f_{avg}$&$f_{best}$&$f_{avg}$&$f_{best}$&$f_{avg}$\\ 
\midrule
\text{v62-c124}&84.26&80.70&84.50&80.80&99.27&82.60&98.98&\textbf{94.00}&73.68&54.50&98.49&90.20&\textcolor{blue}{99.48}&89.10\\
\text{v68-c121}&97.14&91.90&\textcolor{blue}{97.74}&89.00&96.29&\textbf{92.20}&86.48&80.90&65.04&45.70&94.07&91.60&97.21&\textbf{92.20}\\
\text{v69-c135}&93.06&88.30&98.31&88.00&\textcolor{blue}{98.45}&88.90&97.75&\textbf{93.50}&89.74&81.30&93.37&89.70&98.31&92.10\\
\text{v75-c286}&94.66&82.50&94.56&81.50&85.61&77.50&94.43&85.20&96.92&66.30&84.68&81.70&\textcolor{blue}{98.46}&\textbf{87.00}\\
\text{v91-c218}&87.69&83.10&90.75&82.40&\textcolor{blue}{96.13}&\textbf{89.10}&96.01&87.90&65.83&59.50&88.16&80.50&92.98&86.20\\
\text{v94-c215}&95.58&94.00&96.58&95.80&\textcolor{blue}{97.42}&95.30&95.37&94.90&91.41&55.30&95.38&93.80&96.56&\textbf{96.00}\\
\text{v100-c258}&88.95&80.80&87.08&83.30&88.12&77.30&89.00&80.90&66.66&56.60&88.67&81.90&\textcolor{blue}{93.63}&\textbf{85.20}\\
\text{v102-c228}&98.31&90.40&\textcolor{blue}{98.75}&88.50&98.29&\textbf{97.60}&98.40&94.50&69.92&51.30&97.66&90.80&98.55&90.70\\
\text{v103-c179}&90.88&86.40&\textcolor{blue}{97.44}&90.60&95.21&\textbf{92.60}&94.89&92.20&75.27&69.00&82.95&81.90&95.35&90.80\\
\text{v114-c266}&92.05&90.00&94.47&93.40&94.91&\textbf{94.40}&93.01&92.20&80.46&60.10&90.75&89.70&\textcolor{blue}{95.13}&93.70\\
\text{v118-c246}&96.80&96.70&\textcolor{blue}{97.94}&96.20&97.41&97.10&96.64&93.60&90.43&70.50&92.84&87.50&97.72&\textbf{97.50}\\
\text{v129-c300}&97.12&88.70&92.32&90.30&96.57&\textbf{95.00}&97.06&93.80&71.25&55.40&81.41&79.30&\textcolor{blue}{97.22}&90.80\\
\text{v148-c279}&72.76&65.10&82.81&80.90&89.80&83.00&88.10&76.80&57.93&49.30&76.50&71.70&\textcolor{blue}{92.56}&\textbf{87.70}\\
\text{v166-c312}&88.26&86.90&92.27&89.20&\textcolor{blue}{96.11}&89.30&95.94&87.60&79.36&66.20&94.64&89.90&94.01&\textbf{90.60}\\
\text{v167-c323}&85.64&78.90&79.35&77.10&87.37&80.90&\textcolor{blue}{93.53}&81.80&75.55&63.00&87.48&84.70&92.78&\textbf{90.40}\\
\text{v179-c334}&88.89&86.60&84.80&80.30&\textcolor{blue}{94.40}&\textbf{89.10}&93.65&84.70&56.72&54.40&83.10&80.30&92.78&86.70\\
\text{v184-c351}&76.95&66.90&84.39&75.30&92.95&\textbf{87.20}&88.16&74.40&71.16&55.20&81.17&76.60&\textcolor{blue}{94.29}&85.70\\
\text{v187-c399}&85.65&83.30&94.59&85.60&86.10&82.30&\textcolor{blue}{95.92}&87.90&72.38&56.00&79.16&76.40&95.54&\textbf{93.10}\\
\text{v193-c393}&87.28&85.50&89.38&85.70&88.84&88.50&89.53&88.10&66.07&63.00&80.94&79.20&\textcolor{blue}{93.55}&\textbf{90.80}\\
\text{v206-c518}&90.42&86.20&93.53&85.00&86.97&82.50&84.99&82.90&74.36&63.20&80.46&75.50&\textcolor{blue}{96.91}&\textbf{87.90}\\
\text{v224-c523}&83.48&81.20&90.62&87.30&85.88&84.20&89.15&80.50&67.46&53.40&75.45&72.60&\textcolor{blue}{95.29}&\textbf{91.20}\\
\text{v243-c548}&84.33&80.20&87.09&85.30&86.24&83.80&\textcolor{blue}{87.83}&84.10&56.73&52.10&83.53&81.80&86.75&\textbf{86.10}\\
\text{v252-c463}&87.57&77.20&83.01&77.10&80.11&76.50&81.79&77.40&65.29&55.00&73.75&70.90&\textcolor{blue}{94.19}&\textbf{83.90}\\
\text{v256-c572}&86.35&82.20&87.67&83.90&85.17&83.50&\textcolor{blue}{96.20}&\textbf{89.40}&83.71&63.00&84.77&81.50&88.61&85.00\\
\text{v265-c631}&76.33&74.90&77.13&74.00&\textcolor{blue}{86.81}&\textbf{84.10}&80.16&76.80&59.09&45.50&78.17&76.90&86.44&77.90\\
\text{v269-c618}&75.79&72.90&77.28&75.60&77.91&76.60&74.00&69.90&60.31&52.00&72.51&69.70&\textcolor{blue}{83.83}&\textbf{79.80}\\
\text{v284-c632}&78.45&75.30&87.17&76.10&86.35&\textbf{81.10}&\textcolor{blue}{87.83}&80.90&69.88&56.70&75.12&73.50&87.29&78.80\\
\text{v321-c662}&84.74&82.10&88.88&82.40&\textcolor{blue}{90.17}&84.50&89.62&\textbf{86.70}&72.09&65.90&76.73&74.00&88.20&83.50\\
\text{v339-c892}&82.29&76.40&82.99&80.50&84.40&77.40&84.75&79.50&63.90&61.30&70.40&66.60&\textcolor{blue}{87.68}&\textbf{81.30}\\
\text{v1341-c2632}&68.84&68.00&73.50&70.10&\textcolor{blue}{75.52}&\textbf{72.60}&69.91&69.20&61.57&58.40&62.94&61.90&71.09&70.20\\
\midrule
\#best &0&0&4&0&8&11&5&4&0&0&0&0&13&16\\
		\bottomrule
	\end{tabular}
	\caption{The vulnerability coverage percentage (\%) of all versions in binomial distribution scenario. The numbers in bold indicate the best average results, and the numbers in blue represent the best known results.}
	\label{tab:Binomial}
\end{table*}
\setlength{\abovecaptionskip}{20pt}
\begin{table*}[!tb]
	\centering
        \footnotesize
        %\scriptsize
	\begin{tabular}{lrrrrrrrrrrrrrr}
		\toprule
		\multirow{2}*{Instance}&\multicolumn{2}{c}{\AlgNameAbb{0}}&\multicolumn{2}{c}{\AlgNameAbb{1}}&\multicolumn{2}{c}{\AlgNameAbb{2}}&\multicolumn{2}{c}{\AlgNameAbb{3}}&\multicolumn{2}{c}{\AlgNameAbb{4}}&\multicolumn{2}{c}{WalkSAT-BO}&\multicolumn{2}{c}{\AlgNameAbb{}}\\
		\cmidrule(lr){2-3}		\cmidrule(lr){4-5}		\cmidrule(lr){6-7}		\cmidrule(lr){8-9}		\cmidrule(lr){10-11}		\cmidrule(lr){12-13}    \cmidrule(lr){14-15}
		~&$f_{best}$&$f_{avg}$&$f_{best}$&$f_{avg}$&$f_{best}$&$f_{avg}$&$f_{best}$&$f_{avg}$&$f_{best}$&$f_{avg}$&$f_{best}$&$f_{avg}$&$f_{best}$&$f_{avg}$\\ 
\midrule
\text{v62-c124}&\textcolor{blue}{99.99}&89.00&\textcolor{blue}{99.99}&89.00&\textcolor{blue}{99.99}&89.00&83.43&83.40&\textcolor{blue}{99.99}&83.30&83.42&77.90&\textcolor{blue}{99.99}&\textbf{94.50}\\
\text{v68-c121}&99.99&95.20&99.99&\textbf{100.00}&\textcolor{blue}{100.00}&\textbf{100.00}&99.99&\textbf{100.00}&57.17&47.70&85.62&76.20&99.99&\textbf{100.00}\\
\text{v69-c135}&\textcolor{blue}{99.99}&\textbf{100.00}&\textcolor{blue}{99.99}&\textbf{100.00}&\textcolor{blue}{99.99}&\textbf{100.00}&\textcolor{blue}{99.99}&\textbf{100.00}&71.21&57.10&85.80&81.00&\textcolor{blue}{99.99}&\textbf{100.00}\\
\text{v75-c286}&\textcolor{blue}{99.99}&\textbf{100.00}&\textcolor{blue}{99.99}&\textbf{100.00}&\textcolor{blue}{99.99}&\textbf{100.00}&\textcolor{blue}{99.99}&\textbf{100.00}&85.79&76.20&99.98&95.20&\textcolor{blue}{99.99}&\textbf{100.00}\\
\text{v91-c218}&\textcolor{blue}{99.99}&90.40&\textcolor{blue}{99.99}&95.20&\textcolor{blue}{99.99}&\textbf{100.00}&\textcolor{blue}{99.99}&95.20&57.23&52.30&71.48&66.70&\textcolor{blue}{99.99}&\textbf{100.00}\\
\text{v94-c215}&99.98&90.50&85.78&85.80&\textcolor{blue}{99.99}&\textbf{95.20}&85.78&85.80&85.72&66.70&85.71&76.20&\textcolor{blue}{99.99}&\textbf{95.20}\\
\text{v100-c258}&\textcolor{blue}{99.99}&\textbf{100.00}&\textcolor{blue}{99.99}&\textbf{100.00}&\textcolor{blue}{99.99}&\textbf{100.00}&\textcolor{blue}{99.99}&\textbf{100.00}&57.26&47.60&99.98&81.00&\textcolor{blue}{99.99}&\textbf{100.00}\\
\text{v102-c228}&99.98&\textbf{100.00}&\textcolor{blue}{99.99}&\textbf{100.00}&\textcolor{blue}{99.99}&\textbf{100.00}&99.98&\textbf{100.00}&57.34&47.80&99.98&95.20&\textcolor{blue}{99.99}&\textbf{100.00}\\
\text{v103-c179}&\textcolor{blue}{99.99}&\textbf{100.00}&\textcolor{blue}{99.99}&\textbf{100.00}&\textcolor{blue}{99.99}&\textbf{100.00}&\textcolor{blue}{99.99}&\textbf{100.00}&71.40&61.80&99.98&90.50&\textcolor{blue}{99.99}&\textbf{100.00}\\
\text{v114-c266}&\textcolor{blue}{99.98}&\textbf{100.00}&\textcolor{blue}{99.98}&\textbf{100.00}&\textcolor{blue}{99.98}&95.20&\textcolor{blue}{99.98}&\textbf{100.00}&57.05&47.60&71.37&71.40&\textcolor{blue}{99.98}&\textbf{100.00}\\
\text{v118-c246}&99.98&\textbf{100.00}&\textcolor{blue}{99.99}&\textbf{100.00}&\textcolor{blue}{99.99}&\textbf{100.00}&\textcolor{blue}{99.99}&\textbf{100.00}&85.66&57.10&71.67&66.90&\textcolor{blue}{99.99}&\textbf{100.00}\\
\text{v129-c300}&99.98&\textbf{100.00}&\textcolor{blue}{99.99}&\textbf{100.00}&\textcolor{blue}{99.99}&\textbf{100.00}&\textcolor{blue}{99.99}&\textbf{100.00}&74.96&50.00&99.98&83.40&\textcolor{blue}{99.99}&\textbf{100.00}\\
\text{v148-c279}&\textcolor{blue}{99.98}&\textbf{100.00}&\textcolor{blue}{99.98}&\textbf{100.00}&\textcolor{blue}{99.98}&\textbf{100.00}&\textcolor{blue}{99.98}&\textbf{100.00}&62.40&50.00&87.58&87.60&\textcolor{blue}{99.98}&\textbf{100.00}\\
\text{v166-c312}&\textcolor{blue}{99.98}&\textbf{100.00}&\textcolor{blue}{99.98}&\textbf{100.00}&\textcolor{blue}{99.98}&\textbf{100.00}&\textcolor{blue}{99.98}&\textbf{100.00}&87.53&70.90&87.53&83.40&\textcolor{blue}{99.98}&\textbf{100.00}\\
\text{v167-c323}&\textcolor{blue}{99.98}&\textbf{100.00}&\textcolor{blue}{99.98}&\textbf{100.00}&\textcolor{blue}{99.98}&\textbf{100.00}&\textcolor{blue}{99.98}&\textbf{100.00}&74.99&58.30&87.47&79.20&\textcolor{blue}{99.98}&\textbf{100.00}\\
\text{v179-c334}&99.97&\textbf{100.00}&\textcolor{blue}{99.98}&\textbf{100.00}&\textcolor{blue}{99.98}&\textbf{100.00}&99.97&\textbf{100.00}&62.67&50.10&99.96&87.50&\textcolor{blue}{99.98}&\textbf{100.00}\\
\text{v184-c351}&\textcolor{blue}{99.97}&\textbf{100.00}&\textcolor{blue}{99.97}&\textbf{100.00}&\textcolor{blue}{99.97}&\textbf{100.00}&\textcolor{blue}{99.97}&\textbf{100.00}&87.54&62.50&75.02&70.90&\textcolor{blue}{99.97}&\textbf{100.00}\\
\text{v187-c399}&99.97&\textbf{100.00}&99.97&\textbf{100.00}&\textcolor{blue}{99.98}&\textbf{100.00}&\textcolor{blue}{99.98}&\textbf{100.00}&74.92&54.10&99.96&91.60&\textcolor{blue}{99.98}&\textbf{100.00}\\
\text{v193-c393}&\textcolor{blue}{99.98}&\textbf{100.00}&\textcolor{blue}{99.98}&\textbf{100.00}&\textcolor{blue}{99.98}&\textbf{100.00}&99.97&\textbf{100.00}&75.09&45.90&87.54&83.30&\textcolor{blue}{99.98}&\textbf{100.00}\\
\text{v206-c518}&99.96&\textbf{100.00}&\textcolor{blue}{99.97}&\textbf{100.00}&\textcolor{blue}{99.97}&\textbf{100.00}&\textcolor{blue}{99.97}&\textbf{100.00}&50.07&37.60&99.96&95.80&99.96&\textbf{100.00}\\
\text{v224-c523}&99.96&\textbf{100.00}&\textcolor{blue}{99.97}&\textbf{100.00}&\textcolor{blue}{99.97}&\textbf{100.00}&99.96&\textbf{100.00}&62.41&58.30&75.04&74.90&99.96&\textbf{100.00}\\
\text{v243-c548}&99.96&\textbf{100.00}&99.96&\textbf{100.00}&99.96&\textbf{100.00}&99.96&\textbf{100.00}&62.62&54.20&99.95&83.40&\textcolor{blue}{99.97}&\textbf{100.00}\\
\text{v252-c463}&99.96&\textbf{100.00}&\textcolor{blue}{99.97}&\textbf{100.00}&\textcolor{blue}{99.97}&\textbf{100.00}&99.96&\textbf{100.00}&62.49&50.00&87.54&79.20&\textcolor{blue}{99.97}&\textbf{100.00}\\
\text{v256-c572}&\textcolor{blue}{99.97}&\textbf{100.00}&\textcolor{blue}{99.97}&\textbf{100.00}&\textcolor{blue}{99.97}&\textbf{100.00}&\textcolor{blue}{99.97}&\textbf{100.00}&44.51&33.40&88.82&85.20&\textcolor{blue}{99.97}&\textbf{100.00}\\
\text{v265-c631}&99.96&\textbf{100.00}&\textcolor{blue}{99.97}&\textbf{100.00}&\textcolor{blue}{99.97}&\textbf{100.00}&99.96&\textbf{100.00}&66.56&44.40&99.95&85.20&\textcolor{blue}{99.97}&\textbf{100.00}\\
\text{v269-c618}&\textcolor{blue}{88.94}&\textbf{88.90}&\textcolor{blue}{88.94}&\textbf{88.90}&\textcolor{blue}{88.94}&\textbf{88.90}&\textcolor{blue}{88.94}&\textbf{88.90}&55.59&51.80&88.81&81.40&\textcolor{blue}{88.94}&\textbf{88.90}\\
\text{v284-c632}&99.96&\textbf{100.00}&99.96&\textbf{100.00}&99.96&\textbf{100.00}&99.96&\textbf{100.00}&77.78&70.40&77.83&77.70&\textcolor{blue}{99.97}&\textbf{100.00}\\
\text{v321-c662}&\textcolor{blue}{99.96}&\textbf{100.00}&\textcolor{blue}{99.96}&\textbf{100.00}&\textcolor{blue}{99.96}&\textbf{100.00}&\textcolor{blue}{99.96}&\textbf{100.00}&66.70&59.30&88.88&85.20&\textcolor{blue}{99.96}&\textbf{100.00}\\
\text{v339-c892}&99.95&99.90&99.95&\textbf{100.00}&\textcolor{blue}{99.96}&\textbf{100.00}&99.95&\textbf{100.00}&77.79&63.00&88.79&81.50&99.95&\textbf{100.00}\\
\text{v1341-c2632}&99.80&\textbf{99.80}&\textcolor{blue}{99.81}&\textbf{99.80}&\textcolor{blue}{99.81}&\textbf{99.80}&\textcolor{blue}{99.81}&\textbf{99.80}&72.80&66.70&81.72&78.70&\textcolor{blue}{99.81}&\textbf{99.80}\\
\midrule
\#best &15&25&24&27&28&28&18&27&1&0&0&0&26&30\\
  \bottomrule
	\end{tabular}
	\caption{The vulnerability coverage percentage (\%) of all versions in power-law distribution scenario. The numbers in bold indicate the best average results, and the numbers in blue represent the best known results.}
	\label{tab:Power-Law}
\end{table*}

\newpage
\bibliographystyle{ACM-Reference-Format}
\bibliography{sample}

\end{document}

%% file: lab-header.tex
\usepackage{lab}
\usepackage{float}
\newcommand{\yw}[1]{#1}

\newcommand{\xsec}[1]{\section{\xtitle{#1}}}
\newcommand{\xsubsec}[1]{\subsection{\xtitle{#1}}}
\newcommand{\xsubsubsec}[1]{\subsubsection{#1}}

\newcommand{\AlgName}{SAT-based Bayesian optimization}
\newcommand{\AlgNameAbb}{SAT-BO}

\newcommand{\RandBack}{Random backtracking}
\newcommand{\randBack}{random backtracking}
\newcommand{\RandFilpCandid}{Random flipping of decision variables}
\newcommand{\randFilpCandid}{random flipping of decision variables}
\newcommand{\RandFilpIndepend}{Random flipping of independent variables}
\newcommand{\randFilpIndepend}{random flipping of independent variables}
\newcommand{\filpStrageName}{variable flipping strategies}
\newcommand{\verificationRule}{verification rule}

\newcommand{\formula}{F}

\newcommand{\BestSolution}{s^*}
\newcommand{\sampleSet}{S}
\newcommand{\sampleValue}{f}
\newcommand{\prefer}{\delta}
\newcommand{\elitePool}{P}

\newcommand{\determin}{D}
\newcommand{\unAssVarSet}{U}

\newcommand{\maxSolNum}{maxNum}
\newcommand{\prelevel}{p}

\newcommand{\recoveryRate}{\theta}
\newcommand{\randBackProbability}{\rho}
\newcommand{\Equation}[1]{Eq{#1}.}
\newcommand{\Fig}{Figure}
\newcommand{\Tab}{Table}
\newcommand{\Baye}{Bayesian}
\usepackage{color}
\if@szxdraft@
	\RequirePackage{ulem}
	\newcommand{\zqyAdd}[1]{\textcolor{blue}{#1}}
	\newcommand{\zqyDel}[1]{\textcolor{gray}{\sout{#1}}}
\else
	\newcommand{\zqyAdd}[1]{{#1}}
	\newcommand{\zqyDel}[1]{}
\fi

%% file: lab-body.tex
\xsec{Introduction}\label{sec:intro}

\xsubsec{Background}\label{sec:background}

% \begin{table*}
%     \centering
%     \footnote{}
%     \begin{tabular}{ccc}
%         \hline
%         中文 & 曾用 & 修改  \\
%         \midrule
%         核验规则& verification rule / validation rule& verification rule(\textit{verificationRule{}})\\
%         场景覆盖度 &scenario coverage& traffic coverage\\
%         场景覆盖分量&vulnerability coverage component&\\
%         精英解池& elite solution pool&\\
%         分配偏好&assignment preference&\\
%         基于SAT的贝叶斯优化& SAT-Based Bayesian Optimization (SBBO) & SAT-Based Bayesian Optimization (SAT-BO) \\
%         \hline
%     \end{tabular}
%     \caption{proper noun}
%     \label{tab:my_label}
% \end{table*}

% \szxNote{Unify representation: "satisfied request" passes rules; "unsatisfied request" is blocked by rules; "filter" means select requests where the combination of conditions are matched.
% Unmatch -> unsatisfied,
% match -> satisfied,
% filtering rule -> validation rule}
%}
% \xoutline{The problem description in simple words.}

\par
% 数据安全至关重要，因此添加过滤规则以对流量进行校验是确保数据安全的一种重要方式。这些规则可以用来识别三种情况下的流量：未命中规则的流量、命中规则的正常流量以及命中规则的异常流量。

%%%%%%%%%%%%%%%%%%%%%
%%%%【WZ】
%%%% 介绍基本概念%%%%%%%%%%%%%%%%%%%%%
%%%% 
% 问题个人理解的备注：核验规则编码成SAT算例 --> 流量篡改字段field1,从而生成一组篡改流量，评价SAT解覆盖的流量，找到覆盖流量最大的SAT解：S1 --> 不断与BO交互，直到找到最大的流量覆盖SAT解Smax --> 将Smax decode后的核验规则对应的field1字段取反得到新的核验规则
% 所以这里的Tamper Action事实上指的是流量中篡改的字段

% %%%%%%%%%% Text
\begin{figure*}[htbp]
\centering
\includegraphics[scale=0.25]{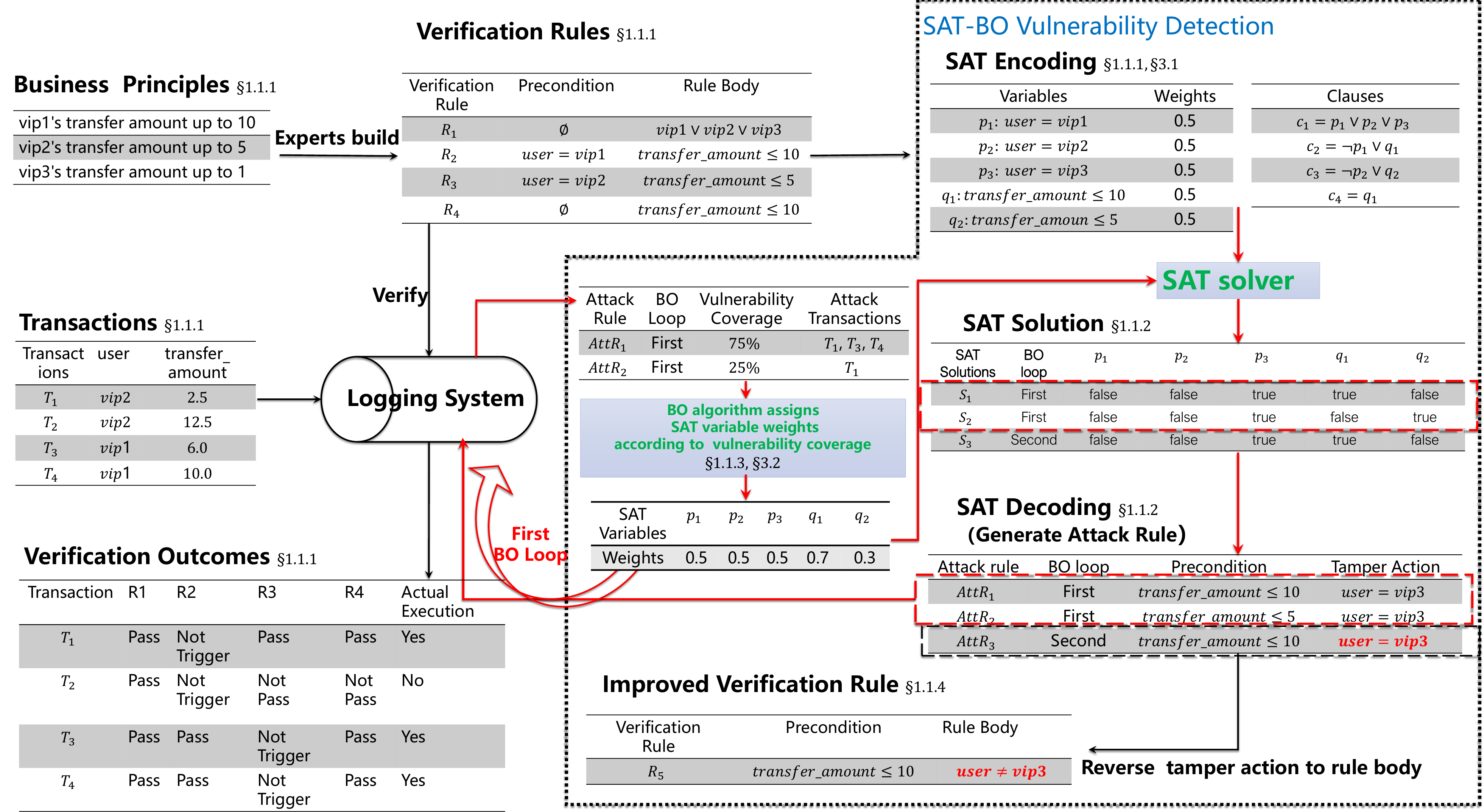}
% \caption{We plot the verification and vulnerability detection systems. The verification rule derived from buiness principles are logic expression consisting precondition and rule body. For example R2 is $IF$ user=vip1 $THEN$ transfer\_amount$\leq$10. It is straightfard that the verification rules lack check of transactions of vip3, which is a vunerability. These verification rules are applied to each transaction to ascertain compliance with both the precondition and the rule body in verification system, and achieve the outcomes categorized as "Pass", "Not Pass", or "Not Trigger". In the vulnerability detection system, SAT-BO method firstly encode the precondition and rule body of verification rules as variables of SAT problems. Then formulate the tampering transaction process as SAT problem and derive SAT solutions. Then the SAT solutions is converted to attack rule with the given tampered field "user". Then the attack rules interact with transaction logging system to see how many transactions trigger the attack rules, i.e., get vulnerability coverage. BO algorithm utilizes these vulnerability coverage to assign weights to the SAT variables. After the BO loop ends, we obtain the attack rule with the largest vulnerability coverage, i.e., $IF$ transfer\_amount$\leq$10 $THEN$ tamper user=vip1. the implication is that transactions fulfilling the precondition transfer\_amount $\leq$ 10 will consequently have the "user" field altered to vip3. Hence, we get new verification rule by reversing the tampered body of attack rule, i.e., from "user=vip3" to "user$\neq$vip3 }
\caption{ We present the structures of both the verification and vulnerability detection systems. The verification system utilizes rules framed as logical expressions with preconditions and rule bodies, based on business principles. In this example, "user" and "transfer\_amount" are fields representing user’s authorization level and the
intended transfer amount (in thousands of dollars), respectively. It is straightforward that the verification rules lack check of transactions of vip3, which is a vulnerability. These rules are systematically applied to transactions, and results are classified as "Pass", "Not Pass" or "Not Trigger". Within the vulnerability detection system, the SAT-BO method encodes verification rule components into SAT variables. It then models verification rules as a SAT problem and obtains solutions that inform the construction of attack rules for the "user" field. These rules are evaluated via the transaction logging system to measure vulnerability coverage. The Bayesian optimization algorithm utilizes this coverage to weight SAT variables. The iterative BO process culminates in an attack rule with maximum coverage: $IF$ transfer\_amount $\leq$ 10 $THEN$ tamper user $=$ vip3. Hence, we get a new verification rule by reversing the tampered action of attack rule, i.e., from "user = vip3" to "user $\neq$ vip3".}
\label{businessFigure}
\end{figure*}

The detailed records of transactions are maintained in logging system to ensure traceability and security. To mitigate the risk of financial losses, it is imperative for these transactions to adhere stringently to established {\bf{business principles}}{\footnote{The key technical terms used in this paper are highlighted in bold and their corresponding meanings are summarized in Table \ref{tab:terms}.}} derived from empirical industry experience. In conventional practice, experts develop a set of logical expressions known as {\bf{verification rules}} based on these principles to scrutinize transactions, thereby identifying and flagging any potentially fraudulent activity. Typically, verification rules are crafted manually by experts, which may not comprehensively encapsulate the intended business principles, thereby introducing potential {\bf{vulnerabilities}}, which are the transactions that violate business principles but not detected by the verification rules. It is therefore essential to evaluate the presence of these vulnerabilities within the verification rules, i.e., {\bf{vulnerability detection}}, and improve the verification rules accordingly. In this paper, we present a framework named SAT-BO designed to detect vulnerabilities by {\bf{attacking}} the verification rules. Our approach leverages Satisfiability (SAT) modeling of the verification rules and employs Bayesian optimization (BO) to refine the detected vulnerabilities. In the following, we first elucidate the fundamental concepts with a practical example, then describe the motivation and provide an overview of SAT-BO.
% Usually, verification rules are created manually by experts, which may not fully capture the intended business principles, leading to potential vulnerabilities. Thus, it is crucial to assess the presence of potential vulnerabilities within these rules—known as vulnerability detection—and improve the verification rules accordingly. In this paper, we propose a framework called SAT-BO to detect vulnerabilities by attacking the verification rules. Our approach utilizes Satisfiability (SAT) modeling of the verification rules and employs Bayesian Optimization (BO) to further refine the detection of vulnerabilities. In the following, we firstly demonstrate more details of the basic concepts with an pratical example then introduce the motivation and overview of SAT-BO.
\xsubsubsec{Business principles and verification rules}

As illustrated in \Fig~\ref{businessFigure}, a bank transfer system has implemented three distinct authorization tiers for users, designated as vip1, vip2, and vip3. Each tier is associated with specific transfer limits: vip1 is allowed up to 10,000 dollars, vip2 up to 5,000 dollars, and vip3 up to 1,000 dollars per transaction. These tiers and limits reflect the underlying business principle governing the system. Consider the following transactions: $T_1$: user = vip2, transfer\_amount = 2.5; $T_2$: user = vip2, transfer\_amount = 12.5; $T_3$: user = vip1, transfer\_amount = 6.0; $T_4$: user = vip1, transfer\_amount = 10.0. Here, "user" and "transfer\_amount" are key {\bf{business fields}} that denote the user's authorization level and the intended transfer amount (in thousands of dollars), respectively. 

According to the business principles, it is postulated that experts would formulate verification rules comprising preconditions and rule bodies. For instance in \Fig~\ref{businessFigure}, the precondition of verification rule $R_3$ stipulates user = vip2, while its rule body mandates transfer\_amount $\leq$ 5. This can be represented by the logical expression: $IF$ user = vip2 $THEN$ transfer\_amount $\leq$ 5. These verification rules are applied to each transaction to ascertain compliance with both the preconditions and the rule bodies in verification system. The outcomes of verification process are categorized as "Pass", "Not Pass", or "Not Trigger". "Pass" indicates that the transaction fulfills both the precondition and the rule body; "Not Pass" indicates that the transaction fulfills the precondition but violates the rule body; "Not Trigger" indicates that the transaction does not meet the precondition. Hence, in the \Fig~\ref{businessFigure}, 
the verification of the transactions $T_1$ (user = vip2, transfer\_amount = 2.5) and $T_2$ (user = vip2, transfer\_amount = 12.5) yields the following results
 transaction $T_1$ results in "Pass", "Not Trigger", "Pass", and "Pass" for the four rules respectively. 
 Transaction $T_2$ results in "Pass", "Not Trigger", "Not Pass", and "Not Pass". Any transaction receiving a "Not Pass" outcome is consequently invalidated and not executed.
%%%%%%%%%%%%%%%%%%%%%%%%%%%%%%%%%%%%%%%%%%%%%%%%%%%%%%%%%%%%%%

%%%%%%%%%介绍如何检测漏洞
%如果核对规则没有完全反映业务准则，不触发/通过核对规则的业务数据可能会违反业务准则，我们把这种情况叫做核对规则出现了安全漏洞，所以检测核对规则是否存在潜在安全漏洞至关重要。%如图1中所示，如果业务数据为 user=vip3，transfer-amount=10，那么该流量能够通过所有核对规则，但它不符合业务准则。通常的检测手段是通过构造”虚假“的业务数据，触发核对规则对该数据进行核验，通过核对规则的核验结果判断”虚假“的业务数据是否为安全漏洞，这样的检测手段又叫做“漏洞攻击”（vulnerability detection），对于”虚假“业务数据的一次核验称为一次攻击。同时，构造的”虚假“业务数据的方式需要具备如下特性：1）”虚假“的业务数据需要和真实业务数据尽可能相同；如果”虚假“的业务数据和真实业务数据差距过大，那么“虚假”业务数据就不具备业务意义，所以这样的漏洞攻击是无效的。那么现有的解决办法是直观的：只修改真实业务数据的一个业务字段，该业务字段通常由业务专家指定，我们成该业务字段为 篡改字段。后续的方案设计也延续这种单篡改字段的攻击方式；2）”虚假“的业务数据不能和历史真实业务数据完全相同；我们假设已有核对规则对于历史发生的业务数据是没有漏洞的，如果出现过因核对规则漏洞导致的资金损失，会做为重大业务故障并快速修复核对规则；3) 核对规则对于”虚假“的业务数据的核验结果是未触发或通过，这说明核对规则没有发现这一次攻击，即”虚假“的业务数据“绕过”了核对规则，是一个潜在的漏洞；4）因为核对规则的执行会消耗大量计算资源，所以需要控制”虚假“业务数据的数量；
% %%%%%%%%%% Text

\xsubsubsec{Vulnerability Detection and Attack}
In instances where verification rules do not fully encapsulate business principles, the transactions with verification outcomes of "not trigger" or “pass” may violate the business principles. Such transactions are referred to as {\bf{vulnerabilities}} within the verification rules. Consequently, it is crucial to assess the presence of potential vulnerabilities within these rules.
As depicted in \Fig~\ref{businessFigure}, it is clearly that the existing verification rules lack of check for vip3 and the transactions of vip3 will be not secure. If we consider a transaction indicating user = vip3 and transfer\_amount = 10, although this transaction may pass all verification rules, it contravenes the business principles. 

To detect such inconsistencies, a widely utilized strategy entails the creation of synthetic transactions that are presumed to be vulnerabilities. This detection method is known as {\bf{vulnerability detection}}. Each verification of synthetic transaction represents an {\bf{attack}} in this context. The constructions of synthetic transactions targeting vulnerabilities should exhibit the following characteristics: 
(1) The synthetic transactions must closely resemble actual transactions. If the disparity between synthetic and real transactions is substantial, the relevancy of the synthetic transactions is lost, rendering the vulnerability attack ineffective. Hence, a straightforward solution is to modify only one business field of real transaction, which is typically designated by experts. This field is referred to as the {\bf{tampered field}}. Subsequent strategies also employ this {\bf{single-field tampering approach}}; (2) The synthetic transactions must not be identical to any historical transaction. It is assumed that existing verification rules are robust against historical transactions; any financial losses due to vulnerabilities would have been identified as significant operational failures and the verification rules would have been promptly amended; (3) If the verification rules yield outcomes of "Not Trigger" or "Pass" when applied to synthetic transactions, it suggests that the rules failed to detect the attack, indicating a potential vulnerability.
 \\

%%%%%%%%%%%%%%%%%整篇文章的动机和主要贡献 
% Text
\xsubsubsec{SAT-BO Motivation and Contributions}
Current methods involve generating synthetic transactions by randomly selecting values within the range of the given tampered field with the intention of producing transactions that constitute vulnerabilities. This approach encounters several issues: Firstly, there is no guarantee that the synthetic transactions will not trigger or pass the verification rules, potentially resulting in ineffective attacks and wastage of computational resources allocated for rule execution; Moreover, the synthetic transaction alone cannot be used to directly formulate new verification rules that would address and close the identified vulnerabilities. 

To resolve the identified challenges, we conceptualize the generation vulnerability through SAT modeling. 
The application of SAT problem-solving ensures that synthetic transactions are formulated in such a manner that they neither inadvertently activate nor successfully pass existing verification rules. %这句话不太理解####
Within our framework, the procedure to create these transactions is defined as the {\bf{attack rule}}, constituted by a precondition and a tamper action. 
Consider the attack rule 1 (abbr. $AttR_1$) in \Fig~\ref{businessFigure} as an illustrative example, where the tampered field is designated as "user". The precondition and the tampered action correspond to transfer\_amount $\leq$ 10 and user = vip3, the implication is that transactions fulfilling the precondition transfer\_amount $\leq$ 10 will consequently have the "user" field altered to vip3, as encapsulated by the construct: $IF$ transfer\_amount $\leq$ 10 $THEN$ tamper user = vip3. 
Meanwhile, the solutions obtained from the SAT problems can be directly converted into novel verification rules, thereby proactively countering identified security vulnerabilities.

Moreover, we have introduced a measure named {\bf{vulnerability coverage}} to assess the significance of the attack rules. This measure quantifies the proportion of transactions that are affected by the specified attack rule. Notwithstanding the usefulness of vulnerability coverage, determining its value requires interactions with the logging system, which incurs cost and is constrained by the number of permissible interactions. The inherent limitation of the SAT modeling is that solutions may be numerous, rendering a comprehensive evaluation of all possible solutions infeasible. To address this, we propose the use of BO algorithm, aimed at maximizing vulnerability coverage. BO algorithm facilitates the efficient direction of the SAT solver in generating attack rules of greater significance within limited interaction loops.

Hence, our main contributions are summarized as follows:
\begin{szxitem}
	\item
 The SAT-BO method effectively encodes the intricate constraint relationships inherent in existing verification rules into SAT formulations to derive attack rules. These rules are designed to precisely target specific vulnerabilities within tampered transactions. Additionally, we have developed a specialized and efficient SAT solver, which is tailored to the characteristics of the problem.
	% SAT-BO encodes complex constraint relationships of existing verification rules into the SAT problems and derive the attack rules ensuring the tampered transactions effectively target specific vulnerabilities. Moreover, we design  a specialized and efficient SAT solver tailored to the characteristics of the problem.

	\item
 To quantitatively assess the impact of these attack rules, we introduce the novel metric of vulnerability coverage. This metric measures the proportion of transactions influenced by a given attack rule, reflecting the rule's significance. Obtaining value of vulnerability coverage is resource-intensive and testing every attack rule derived from SAT problems is impractical. To circumvent these constraints, we employ BO algorithm to refine the process of identifying and prioritizing attack rules that offer the most extensive transaction coverage.
	% We define the novel metric of vulnerability coverage to quantify the proportion of transactions affected by an attack rule, thereby assessing the significance of these rules. Since get vulnerability coverage is costly and we cannot test all the derived attack rules from SAT problems. We introduce Bayesian Optimization (BO) algorithm that streamlines the identification of critical vulnerabilities and prioritizes the creation of attack rules that have the most substantial coverage on the system.
	
	\item
    To enhance the vulnerability coverage of SAT solutions, we optimize from two aspects: diversity and determination of solutions. On one hand, we employ three random strategies including random backtracking, random flipping of decision variables, and random flipping of independent variables to enhance the diversity of SAT solutions. On the other hand, we use BO algorithms to recommend SAT variable assignment weight, guiding the search direction of the SAT solver and thus enhancing the determination of solutions.
\end{szxitem}

\xsubsubsec{SAT-BO Overview}
As demonstrated in \Fig~\ref{businessFigure}, we formulate verification rules as SAT problems, which is delineated in Section~\ref{sec:modeling}. The variables in this problem are represented as $p_1$: user = vip1, $p_2$: user = vip2, $p_3$: user = vip3, $q_1$: transfer\_amount $\leq$ 10, and $q_2$: transfer\_amount $\leq$ 5. Upon solving the SAT problem, we obtain two distinct solutions: 1) $p_1=false$, $p_2=false$, $p_3=true$, $q_1=true$, $q_2=false$; and 2) $p_1=false$, $p_2=false$, $p_3=true$, $q_1=false$, $q_2=true$. With the given tampered field "user", we utilize the "single-field tampering approach" to decode the SAT solutions and derive the corresponding attack rules: $AttR_1$: $IF$ transfer\_amount $\leq$ 10 $THEN$ tamper user = vip3 and $AttR_2$: $IF$ transfer\_amount $\leq$ 5 $THEN$ tamper user = vip3. Subsequently, we evaluate the attack rules to determine the coverage of transactions they impact, i.e., their vulnerability coverage. Given the four transactions, the first attack rule achieves $75\%$ vulnerability coverage, i.e., $T_1$, $T_3$ and $T_4$ trigger attack rule. By comparison, only transaction $T_1$ satisfies the precondition of the second attack rule (transfer\_amount $\leq$ 5) resulting in the vulnerability coverage of $25\%$. Considering the vulnerability coverage of these two attack rules, the BO algorithm will assign greater weight to the variable $q_1$ that maximizes vulnerability coverage. In the subsequent iteration of BO, the SAT solver returns solution $S_3$: $p_1=false$, $p_2=false$, $p_3=true$, $q_1=true$, $q_2=false$ with the variable weights from  the last iteration. Consequently, the final attack rule is established as $IF$ transfer\_amount $\leq$ 10 $THEN$ tamper user = vip3. Reversing the tampered action of this attack rule from "user = vip3" to "user $\neq$ vip3" generates a new verification rule: $IF$ transfer\_amount $\leq$ 10 $THEN$ tamper user $\neq$ vip3. This new rule is implemented to avert the vulnerability where no existing verification rule accounts for transactions by vip3 users.

\xsec{Literature review}\label{sec:literature}
\xsubsec{Fraud Detection}
% fraud detection的方法主要分为两类，利于rule 和 machine learning（ML） algorithms 去 verify transactions。 The need of automatic systems able to detect frauds from historical data led to the design of a number of machine learning algorithms for fraud detection. Supervised methods, typically based on binary classification, as well as unsupervised and one-class classification have been proposed in literature. 然而，ML-based的方法执行效率以及可解释性比较差， verifications rules 主要是由专家编写的逻辑表达式，逻辑表达式的执行效率以及可解释性都是比较高的，所以是online production environment的主流方法。但是现有rule based方法最主要的缺点是没有系统化的编写方法保证 verification rules没有 vunerability。只能在fraud出现后，再编写新的verification rules去堵住这个漏洞。而SAT-BO将rule 的 vunerabilities发现过程建模为SAT problem，为verification rule生成提供了vunerability-free的理论保证。

Fraud detection methods can be broadly categorized into two main approaches: rule-based systems~\cite{islam2024rule} and machine learning (ML) algorithms for transaction verification~\cite{CARCILLO2018182,RODRIGUES2022101207}. Various ML-based approaches have been proposed regarding key challenges, such as the imbalance in class distribution due to the relatively low percentage of fraudulent transactions~\cite{7376606}, the concept drift resulting from changes in the distribution of fraudulent activities over time~\cite{6425489,2523813}, and the need for efficient stream processing~\cite{3657286}.
However, ML-based approaches often suffer from limitations in computational efficiency and interpretability. 

In contrast, verification rules, primarily composed of logical expressions crafted by domain experts, exhibit superior execution efficiency and interpretability. Consequently, rule-based methods remain the predominant approach in the cooperate company's online production environments.
Nevertheless, the primary drawback of existing rule-based methods lies in the absence of a systematic approach to ensure the robustness of verification rules against vulnerabilities. Typically, new verification rules are reactively implemented only after the occurrence of fraudulent activities to address identified vulnerabilities. 

To address this limitation, SAT-BO presents an innovative approach by modeling the process of identifying verification rule vulnerabilities as a SAT problem. This approach provides a theoretical foundation for generating vulnerability-free verification rules, thereby enhancing the proactive capabilities of fraud detection systems.

\xsubsec{SAT Problems and Solvers}
SAT problem is the first proven NP-complete problem~\cite{cook1971complexity}, and the academic community has extensive experience in encoding, solving, and applying SAT problems~\cite{gupta1994efficient, obrenovic2019consolidation}. The algorithms for solving SAT problems are mainly divided into complete algorithms and incomplete algorithms. Among them, incomplete algorithms are mainly represented by local search algorithms, such as \textbf{WalkSAT}~\cite{cai2012configuration}. For complete algorithms, the first introduced one is Davis-Putnam-Logemann-Loveland (\textbf{DPLL}) algorithm in the 1960s~\cite{davis1960computing, davis1962machine}, and later the development and continuous iteration of the conflict-driven clause learning (\textbf{CDCL}) algorithm in the 1990s~\cite{silva2003grasp}, along with related research on solvers such as MiniSat, Glucose, Maple\_LCM, and Kissat~\cite{een2003extensible, audemard2009predicting, luo2017effective, fleury2020cadical}. SAT problem-solving algorithms can now quickly solve large-scale constraint satisfaction problem instances. 

Our method \AlgNameAbb{} employs the DPLL-based SAT solver. 
DPLL is a depth-first tree search algorithm that enhances search efficiency by heuristically selecting appropriate variable assignments to imply more variable assignments. DPLL first performs unit propagation to simplify the CNF formula. The {\bf{unit propagation}} determines the variable assignments to satisfy the formula based on the found unit clauses. Then, it recursively checks if the simplified formula is satisfiable. If the formula becomes empty, then a satisfying solution has been found. If the formula exists an empty clause, indicating a conflict with the current assignment. 
Then, DPLL solver assigns some unassigned variables in the formula using the heuristic strategy to further reduce the formula. Once a satisfying assignment is found or the problem is determined to be unsatisfied, DPLL solver terminates.

Based on our comprehensive analysis from the perspective of SAT formulations, the current verification rules do not entirely conflict with each other. Our objective is to utilize SAT solvers to identify hidden constraints that may lead to financial losses, thereby enhancing the existing verification rules. The SAT-BO method effectively encodes the complex constraint relationships inherent in the current verification rules into SAT formulations, enabling the derivation of potential attack rules.
\xsubsec{Bayesian Optimization}
% In SAT-BO, we employ Bayesian optimization algorithm to refine the process of identifying and prioritizing attack rules that offer the most extensive transaction coverage.
% BO algorithm is one of the most efficient black box optimization (BBO) algorithms. 
% BBO is a rapidly advancing field of optimization with widespread applications in complex systems engineering, energy and the environment, materials design, drug discovery, chemical process synthesis, and computational biology \cite{Bajaj2021}. 
The characteristics of the black-box problem are as follows.
The objective function and constraints may be partially or completely unknown \cite{ALARIE2021100011}; The process of obtaining the objective function value and checking the feasibility of constraints can be time-consuming. For example, a single experiment in hyperparameter optimization may take several hours \cite{audet2006finding}, and the process of airfoil trailing-edge noise reduction can even take several days \cite{marsden2007trailing}.
% Sometimes it is necessary to use simulation methods to obtain the objective function value, and the results obtained may be imprecise \cite{huot2019hybrid}, with significant differences in results even after multiple simulations \cite{chen2018smoothing}.

% Common methods for solving BBO problems include meta-heuristics, direct search, and model-based methods \cite{wortmann2016black}.
% Model-based methods have become a hot topic in solving BBO problems in recent years. 
% These methods use a surrogate model to mimic the objective function of the BBO problem, in order to effectively guide the search. Bayesian optimization (BO) is one of the most common model-based methods \cite{bernardo2009bayesian,lian2023conttune}. 
BO consists of two key components: a probabilistic surrogate model, typically a Gaussian process, which models the objective function values based on historical search information and provides an assessment of the uncertainty in the objective function values to guide the search; and an acquisition function, which, based on the information from the probabilistic surrogate model, determines the inputs for sampling after balancing between exploration and exploitation. There are many solvers based on BO algorithms, such as HEBO~\cite{Cowen-Rivers2022-HEBO}, TuRBO~\cite{eriksson2019scalable}, and GPyOpt~\cite{gpyopt2016}. Taking into account the performance and ease of adaptation, this study ultimately chose TuRBO as the solver for BBO problems.

\xsec{The \AlgName{} algorithm}\label{sec:alg}

\xsubsec{Modeling the Problem}\label{sec:modeling}
% 根据问题定义，核验规则是一组包含$n$个变元和$m$个子句的CNF布尔表达式$F$，
% 如果存在一个真值指派$s_t=\{x_i| x_i=true \lor x_i=false ,\forall 1\le i\le n \}$使得该布尔公式的取值为真则称其为该布尔公式的一个解。
% 每个解$s_t$都是一个潜在的攻击规则。定义布尔公式的所有解的集合是$S$，对于每一个解$s_t\in S$都存在一个函数$f(s_t)$表示该攻击规则的场景覆盖度，本研究的目标是寻找一个解$s_t$使得$f(s_t)$最大，如公式$\ref{target}$所示

Given a set of variables $V = \{x_1, \dots, x_n\}$, a literal is a variable $x\in V$ or its negation $\neg x$, a clause is the disjunction of literals, a unit clause contains exactly one literal, and a conjunctive normal form (CNF) formula is a conjunction of clauses. 
A Boolean formula is said to be a solution, if there exists a truth assignment $s=\{x_i| x_i=true \lor x_i=false,\forall 1\le i\le n \}$ such that the Boolean formula takes on a truth value where all clauses of the CNF are assigned true.
Each solution $s$ represents a potential attack rule.
The set of all solutions for the Boolean formula is denoted as $S$.
The verification rules are logic expressions based on business principles and can be formulated as a set of CNF formula denoted as $F$ containing $n$ variables and $m$ clauses.
Accordingly, if the CNF formula evaluates to a truth value, this indicates that the outcomes of the verification rules correspond to "not trigger" or "pass". Conversely, if the formula does not evaluate to a truth value, the outcomes contain "not pass".

% Hence, if the CNF Boolean formula takes on a truth value, it means that the outcomes of the verifications rules are "not trigger" or "pass". Otherwise, the outcomes contain "not pass".
For each solution $s\in S$, a function $f(s)$ exists representing the vulnerability coverage of that attack rule.
The objective is to find a solution $s$ that maximize $f(s)$, as shown \Equation{}~(\ref{target}).
\begin{align}\label{target}
\max{f(s)},\forall s \in S
\end{align}

\begin{figure}[htbp]
% \vspace{-1.4em}

\centering
\includegraphics[width=0.52\textwidth]{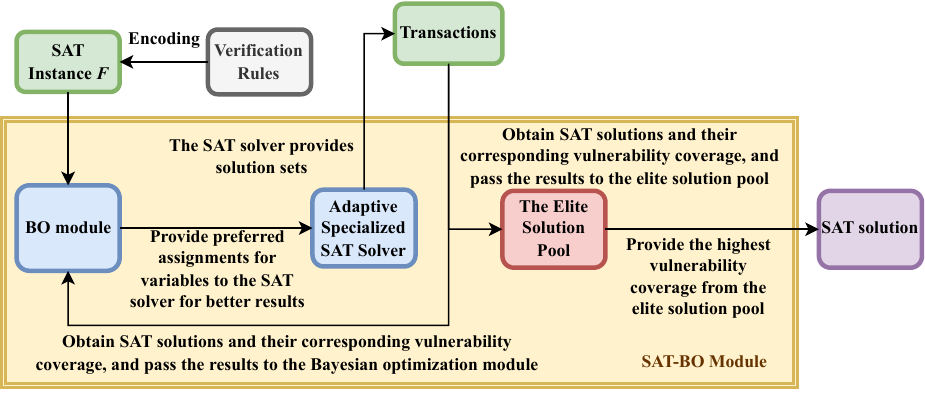}
\caption{Flow chart of the main algorithm framework}
\label{mainFrameFigure}
\vspace{-3.3em}

\end{figure}

Here $f(s)$ represents the vulnerability coverage that can be intercepted when this SAT solution is decoded into attack rules.
Vulnerability coverage quantifies the fraction of transactions that activate the attack rules, i.e., transactions that meet the attack rules' preconditions. Mathematically, it is expressed as: 
\begin{align}\label{vulnerability}
\text{vulnerability coverage} = \frac{\text{$\alpha$}}{\text{\#trans}}
\end{align}
where $\alpha$ represents as number of transactions triggered by attack rule and \#trans represents as total number of transactions.
% The function $f(s)$ is characterized by inherent unpredictability and uncertainty. Furthermore, real-world transactions are often with patterns of fluctuation, periodicity, and mobility, adding complexity to the simulation and explore $f(s)$ particularly important and difficult in this study.

% Specifically, vulnerability coverage is defined as the transactions portion that the attack rules are triggered, i.e., transactions satisfy the preconditions of attack rules. Hence, the
% \begin{align}\label{vulnerability}
% vulnerability coverage = \frac{amount transactions triggered by attack rules }{total transactions amount}
% \end{align}
% Therefore, the function $f(s)$ exhibits characteristics of unpredictability and uncertainty, and due to real-world traffic often exhibits characteristics such as fluctuations, periodicity, and mobility, which makes how to simulate and explore $f(s)$ particularly important and difficult in this study.

\xsubsec{Algorithm Framework of SAT-BO}
\label{sec:framework}

% \zqyAdd{In order to generate attack rules, we transform a series of \verificationRule{s} into SAT instances and propose the \AlgName{} (\AlgNameAbb{}) algorithm to solve the SAT instances.
%find the SAT solution that maximizes the coverage of abnormal traffic by solving the SAT instances.
\AlgNameAbb{} uses an improved SAT solver for the SAT instances, which is a tree search-based SAT solver with completeness, and adopts the \Baye{} optimization method to guide the SAT solver. \zqyDel{improve the process of solving the SAT problem.}
The main flow is shown in \Fig{} $\ref{mainFrameFigure}$.
First, the verification rules are encoded into the SAT instance.
Then, \AlgNameAbb{} solves the problem after coding based on a specialized adaptive DPLL SAT solver and adopts an improved \Baye{} optimization module to guide the search of the SAT solver, where the SAT solver and the \Baye{} optimization module are alternately performed to find a better solution.
Specifically, the \Baye{} optimization module provides the assignment weights of variables to the SAT solver according to corresponding vulnerability coverage, then the SAT solver repairs the sample set based on the received assignment weights, and finally maps the repaired sample set with the corresponding vulnerability coverage and provides it to the \Baye{} optimization module for exploration.
The algorithm gradually approaches the optimal solution, i.e., the SAT solution with the largest vulnerability coverage, by constant interaction between the SAT solver and BO.

% \zqyNote{TODO：\\
%          1. 符号规范化，前后含义一致！！\\
%          2. 符号定义，伪代码涉及到的变量要在正文中作解释\\
%          3. 样本点集->样本集\\
%          4. 命名的规范性，如精英池/精英解池}
         
\begin{algorithm}
    \caption{The main framework of the \AlgName{} algorithm}
    \label{mainFrame}
    \KwIn{Boolean formula $\formula{}$, maximum iterative interaction counter $MaxIter$}
    \KwOut{SAT solution $\BestSolution{}$ with the best vulnerability coverage}
    
    \BlankLine
    % \Function{MAIN}{R,MaxIter}
    %Variable set $\VertexSet \gets init(\formula{})$\\
    %\zqyNote{问题定义部分也要解释说明}\\
    Iteration counter $iter\gets 0$, elite solution pool $\elitePool{}\gets \emptyset$ \label{alg1Line:DataInit1}\\
    
    Sample set of SAT solutions $\sampleSet{} \gets \emptyset$\\
    
    Vulnerability coverage $\sampleValue{}\gets \emptyset$ \label{alg1Line:DataInit2}\\
    
    \While{$iter<MaxIter$\label{alg1Line:frameBegin}}{
         Assignment weights $\prefer{}_i \gets  0.5 ,~\forall i \in V$\\
        $\sampleSet{}\gets adaptiveDPLL(\formula{}, \prefer)\label{alg1Line:frameGetSampleFirst}$ \tcc*[f]{See Algorithm \ref{adaptiveDPLL}}\\

        $\sampleValue{} \gets getSolScenCover
(\sampleSet{})$\label{alg1Line:frameGetSampleEndFirst}\\

        $\elitePool{} \gets \elitePool{}\cup (\sampleSet{}, \sampleValue{})$ \label{alg1Line:frameAddPoolFir}\\
        
        $initCI(\sampleSet{}, \sampleValue{})$\label{alg1Line:initCI}\\
    \While{Confidence interval is sufficiently large and $iter<MaxIter$ \label{alg1Line:iterBegin}}{
        $\prefer{} \gets suggestPrefer()$\label{alg1Line:updatePrefer}\\
        
        $\sampleSet{} \gets adaptiveDPLL(\formula{}, \prefer)$ \label{alg1Line:frameGetSampleSec}\\
        
        $\sampleValue{} \gets getSolScenCover
    (\sampleSet{})$ \label{alg1Line:frameGetSampleEndSec}\\
    
        $\elitePool{} \gets  \elitePool{}\cup(\sampleSet{}, \sampleValue{})$ \label{alg1Line:frameAddPoolSec}\\
        $updateCI(\sampleSet{}, \sampleValue{})$ \label{alg1Line:updateCI}\\
        $iter\gets iter+1$\label{alg1Line:frameEnd}\label{alg1Line:iterEnd}\\
    }
    }
 $\BestSolution \gets$ Find the best solution in $P$\label{alg1Line:frameBestSol}\\
    % \State 
    % \EndFunction
return $\BestSolution$\\
\end{algorithm}

\zqyDel{Specifically, we first transform the \verificationRule{s} into a SAT instance and obtain the maximum iterative interaction counter, which is the number of interactions between the SAT solver and \Baye{} optimization. At the same time, the SAT solver and the \Baye{} optimization module are initialized (lines 1-3).
Then, \AlgNameAbb{} optimizes the solution iteratively by the interaction of the SAT solver and the \Baye{} optimization until the maximum iterative interaction counter is met (lines 5-14).
In each iteration of the \AlgNameAbb, it randomly selects a set of solutions as the sample set $\sampleSet$ from the SAT solver, and calculates the vulnerability coverage $f(\sampleSet)$ for these sample points (lines 6-7). 
The collected samples are placed into an elite solution pool (line 8), and the confidence interval for \Baye{} optimization is initialized (line 9). 
Next, \AlgNameAbb{} iteratively interacts within the confidence interval of the \Baye{} optimization with no more than the maximum iterative interaction counter, otherwise, the algorithm resets the confidence interval of the \Baye{} optimization (lines 10-16).
The \Baye{} optimization module provides multiple sets of assignment weights to the SAT based on historical samples provided by the SAT solver and updates the variable assignment weights of the SAT (lines 11-12). 
The SAT obtains new sample set $\sampleSet$ based on the assignment weights provided by \Baye{} optimization and recalculates the vulnerability coverage $f(\sampleSet)$ (lines 13-14), and then the collected samples are placed into the elite solution pool (line 15). 
Finally, the confidence interval of \Baye{} optimization is updated (line 16). 
Once the specified termination condition is met, \AlgNameAbb{} terminates and returns the subset with the highest vulnerability coverage from the elite solution pool (line 17).
}
\zqyAdd{The main framework of the \AlgNameAbb{} algorithm is described in Algorithm $\ref{mainFrame}$.
%For the SAT instance after the \verificationRule{s} transformation, 
\AlgNameAbb{} first initializes the sample set of solutions $\sampleSet$ and other relevant data (lines \ref{alg1Line:DataInit1}-\ref{alg1Line:DataInit2}).
%we first obtain the maximum iteration counter, which is the number of interactions between the SAT solver and \Baye{} optimization. 
Then, it optimizes the solution iteratively by the interaction of the SAT solver and the \Baye{} optimization until the maximum iteration counter $MaxIter$ is met (lines \ref{alg1Line:frameBegin}-\ref{alg1Line:frameEnd}).
In each iteration, \AlgNameAbb{} first randomly selects a set of solutions as the sample set $\sampleSet$ from the SAT solver, and calculates the vulnerability coverage $f(\sampleSet)$ for these samples (lines \ref{alg1Line:frameGetSampleFirst}-\ref{alg1Line:frameGetSampleEndFirst}). 
%Then, it initializes the confidence interval for \Baye{} optimization (line \ref{alg1Line:initCI}).
Next, \AlgNameAbb{} optimizes the current sample set of solutions $\sampleSet$ by interacting between the SAT solver and the \Baye{} optimization within the confidence interval with no more than the maximum iteration counter. Otherwise, the algorithm resets the confidence interval of the \Baye{} optimization (lines \ref{alg1Line:iterBegin}-\ref{alg1Line:iterEnd}).
The \Baye{} optimization module provides multiple sets of assignment weights to the SAT solver based on historical samples provided by the SAT solver, and then the SAT solver obtains new samples $\sampleSet$ based on the assignment weights provided by \Baye{} optimization and recalculates the vulnerability coverage $f(\sampleSet)$ (lines \ref{alg1Line:updatePrefer}-\ref{alg1Line:frameGetSampleEndSec}).
Finally, the confidence interval of \Baye{} optimization is updated (line \ref{alg1Line:updateCI}). 
Once the specified termination condition is met, \AlgNameAbb{} terminates and returns the SAT solution with the highest vulnerability coverage from the elite solution pool (line \ref{alg1Line:frameBestSol}).
}
\xsubsec{The Customized Adaptive DPLL SAT Solver}\label{sec:init}

\zqyAdd{
% The introduction of the DPLL algorithm framework in the literature review part shows that it is a deterministic algorithm, so the solutions obtained by the DPLL solver are similar.  
Existing DPLL algorithms are deterministic, and the solutions they produce are usually similar~\cite{davis1960computing, davis1962machine}. This is due to the search strategy, where for a given branching variable trace, the DPLL solver must search its subtree before taking another assignment branch for that variable.
Therefore, it may lead to the \Baye{} optimization module only learning the properties of one part of the data and knowing nothing about the properties of other parts of the data, which will make the exploration space become limited. 
To increase the diversity of the solution space and improve the performance of \AlgNameAbb, we introduce three \filpStrageName{} to the DPLL solver.
%Our algorithm mainly uses three random strategies to improve the diversity, robustness, and performance of the model.
}

%These three random strategies aim to introduce variability into the DPLL algorithm, resulting in a broader range of solutions. This diversification is important because it allows for a more comprehensive exploration of the solution space and can potentially lead to better optimization results.
\begin{itemize}
%     \item \zqyNote{随机回溯}\\
% 在算法进行阶段不再单一的回溯到上一层，而是随机回溯到某一层，防止层数比较低的节点长时间不改变。
    \item \textbf{\RandBack{} (RB)}:
During the search process, instead of exclusively backtracking to the immediately preceding level, the algorithm randomly backtracks to a certain level. 
This prevents variables at lower levels from remaining unchanged for a long time.

%The purpose of introducing this random backtracking strategy is to add an element of randomness to the exploration process. In traditional backtracking, the algorithm systematically explores each level, potentially leading to prolonged periods of stagnation, especially at lower levels of the search tree. Random backtracking helps to break this pattern by introducing randomness into the decision-making process.
%By randomly choosing a level to backtrack to, the algorithm has a chance to explore different branches of the search space, potentially uncovering more diverse solutions. This randomness is intended to prevent the algorithm from getting stuck in local optima and to encourage a more dynamic exploration of the solution space. It is a strategy aimed at enhancing the adaptability and diversity of the algorithm's exploration, ultimately contributing to a more effective and robust optimization【自动补充】
%     \item \zqyNote{候选变元的随机翻转}\\
% 在进行候选变元选择时以一定概率随机逆转所取的变元的值。来跳出僵局。防止某个变元长时间不变动。
    \item \textbf{\RandFilpCandid{} (RFDV)}:
The values of the selected decision variables are randomly reversed with a probability. 
This strategy aims to break out of deadlocks and prevent a situation where a particular variable remains unchanged for a long time.
%The key aspect of this strategy is the introduction of probabilistic variable reversal. In traditional DPLL algorithms, the selection and assignment of candidate variables are based on deterministic rules. However, this deterministic approach can sometimes lead the algorithm to get stuck in a particular branch, as it adheres to a fixed path. By introducing probabilistic reversals, the algorithm has the opportunity, under certain probabilities, to deviate from the current path and randomly try another variable assignment. This allows for a more dynamic exploration of the solution space by occasionally breaking free from rigid decision-making patterns.
%The advantage of this strategy is that it helps the algorithm escape potential deadlocks during the search process, where it might otherwise struggle to find better solutions. This is achieved by introducing an element of randomness that allows the algorithm to make more flexible decisions and explore the diversity of the solution space more comprehensively. It assists in avoiding local optima and increases the algorithm's ability to search for better solutions over time.

    % \item \zqyNote{无关变元的随机翻转}\\算法对非决定变元层的变元进行随机赋值。对于不出现在公式$F$中无关紧要的变元进行随机赋值。\textcolor{red}{[句子重复，that is（i.e.）？！]}
    \item \textbf{\RandFilpIndepend{} (RFIV)}:
   The algorithm makes random assignments to non-decision variables, i.e., random assignments to irrelevant variables that do not appear in assignment track during searching $F$. 
\end{itemize}

\zqyAdd{Based on several \filpStrageName, a specialized adaptive DPLL SAT solver is proposed. Algorithm $\ref{adaptiveDPLL}$ shows the main algorithm framework.
It first performs unit propagation 
% $unitPropagation(F,U,D)$ 
 (line \ref{Alg3Line:unitPropagation}), if the formula is empty (lines \ref{Alg3Line:ifBegin}-\ref{Alg3Line:ifEnd}), i.e., a satisfying solution is found, and it is stored to the sample set of solution $\sampleSet$.
If the formula contains an empty clause (lines \ref{Alg3Line:elifBegin}-\ref{Alg3Line:elifEnd}), i.e., the current assignment does not satisfy the clause, then $recoverFToLevel(F,U,D,p)$ heuristically chooses the number of level to return. 
Otherwise, to further simplify the formula, \AlgNameAbb{} uses the \filpStrageName{} to assign values to unassigned variables by $getCandicate(U,\delta)$ (line \ref{Alg3Line:getCandicate}). 

The process is iteratively backtracked until the number of solutions needed $maxNum$ are reached (line \ref{Alg3Line:beginWhile}).
Finally, when the number of solutions found meets the number of solutions needed $maxNum$, the process terminates and returns the sample set $\sampleSet$.
}
\begin{algorithm}
    \caption{adaptiveDPLL$(\formula{}, \prefer)$}
    \label{adaptiveDPLL}
    \KwIn{Boolean formula $\formula{}$, assignment weights for each variable $\delta$} %, number of solutions found $\hasFoundSol$}
    \KwOut{Sample set of solution $\sampleSet$}
   $\determin\gets \emptyset$ \tcc*[f]{$\determin$ is set of assigned variables}\label{Alg3Line:defineD}\\
    $\unAssVarSet\gets [1,n]$ \tcc*[f]{$\unAssVarSet$ is set of unassigned variables}\\
   $p\gets 0$\\
    \While{$|\sampleSet|<maxNum$ }{\label{Alg3Line:beginWhile}
    % $unitPropagation(F,U,D)$ \\
    using unit propagation rule on $F$\label{Alg3Line:unitPropagation}\\
        \uIf{$F=\emptyset$}{\label{Alg3Line:ifBegin}
            $\sampleSet\gets \sampleSet\cup \{s\}$ \tcc*[f]{ $s$ is the SAT solution}\\
            $p\gets getLevel(\sampleSet,U,D,\delta)$\tcc*[f]{See Algorithm \ref{getLevel}}\\
            $recoverFToLevel(F,U,D,p)$ \label{Alg3Line:recoverFToLevel}\\
        }\label{Alg3Line:ifEnd}
        \uElseIf{$F~\text{\rm exists the empty clause}$}{\label{Alg3Line:elifBegin}
            $p\gets getLevel(\sampleSet,U,D,\delta)$\tcc*[f]{See Algorithm \ref{getLevel}}\\
            $recoverFToLevel(F,U,D,p)$\\
        }\label{Alg3Line:elifEnd}
        \Else{\label{Alg3Line:elseBegin}
            % \If{$F=\emptyset$}{
            % $X\gets$ the SAT solution\\
            % $p\gets getLevel(X,U,D,\delta)$\tcc*[f]{See Algorithm \ref{getLevel}}\\
            % $recoverFToLevel(F,U,D,p)$\\
            % continue\\
            % }
            $x\gets getCandicate(U, \delta)$\tcc*[f]{See Algorithm \ref{getCandicate}} \label{Alg3Line:getCandicate}\\
            $D\gets D\cup \{x\}$
        }\label{Alg3Line:elseEnd}
    }\label{Alg3Line:endWhile}
    \BlankLine
    % \hywNote{$\delta$ use to determine whether variable $x$ is $x$ or $\neg x$}
   
\end{algorithm}
% \hywAdd{随后本文具体介绍采用的三种随机策略。对于SAT求解器来说，变元的赋值是布尔变量，但对贝叶斯来说，变元的赋值是连续变量，对于一个变元的赋值偏好，变元的赋值偏好从$0.9$变至$0.91$，对于贝叶斯来说也是有变换的}
% \par 如算法$\ref{getCandicate}$所示，描述了基于自适应的DPLL算法选择候选变元的算法框架。算法在未赋值的变元集合$U$中选择赋值偏好程度比较明显的变元集合$C$，并且根据变元的赋值偏好为集合$c$中的每一个变元进行赋值(lines 3-14).然后算法随机选择一个变元$x$作为候选变元(line 15)。算法根据变元的赋值偏好$prefer_x$，以一定的概率随机将变元的值强行逆转(lines 16-19)。最后返回一个该候选变元对应赋值的文字(line 19)。
\begin{algorithm}
\caption{getCandicate($\unAssVarSet,\prefer$)}
    \label{getCandicate}
    \KwIn{A set of unassigned variables $\unAssVarSet$, assignment weights for each variable $\prefer$}
    \KwOut{Selected variable $x$}
    \BlankLine
    Set of variables with the clearest preference $C\gets \arg\max_{x \in \unAssVarSet} |\prefer_x-0.5|$\\
    \For{$x\in C$}{\label{Alg4Line:whileBegin}
            \If{$\prefer_x<0.5$}{
                $C \gets C \setminus \{ x \} \cup \{\neg x\}$\tcc*[f]{Calibrate negative literals}\\
            }
        }\label{Alg4Line:whileEnd}
        $x\gets$ a randomly picked candidate variable in $C$ \label{Alg4Line:randomCand}\\
        $t\gets$ a randomly picked integer in $[1,100]$ \label{Alg4Line:randomNum}\\
        $\recoveryRate_x\gets \frac{0.5}{|\prefer_x-0.5|}$\label{Alg4Line:GetTheta}\\
        \If{$t\le \recoveryRate_x$}{$x \gets \neg x$ \label{Alg4Line:flipVariable}\\}
        return $x$\\ 
    % \Function{MAIN}{R,MaxIter}
\end{algorithm}
\zqyAdd{Algorithm $\ref{getCandicate}$ shows the $getCandicate(U,\delta)$ framework, which selects decision variables based on the adaptive DPLL algorithm.
It selects the set $C$ of variables with a more pronounced degree of assignment preference among the set $U$ of unassigned variables, and assigns a value to each variable in the set $C$ according to the assignment preference of the variable (lines \ref{Alg4Line:whileBegin}-\ref{Alg4Line:whileEnd}). 
Then, the procedure randomly selects a variable $x$ as a decision variable (line \ref{Alg4Line:randomCand}). 
Next, the procedure reverses the value of a variable with probability $\recoveryRate_x$ according to the assignment preference $\prefer_x$ of the variable $x$ (lines \ref{Alg4Line:randomNum}-\ref{Alg4Line:flipVariable}).  
%Finally, it returns a variable corresponding to the assignment of that candidate variable (line 19).
}

% \par 如算法$\ref{getLevel}$所示，描述了基于自适应的DPLL算法随机回溯层的启发式算法框架。初始化回溯变元层为0(即没有特殊要回到的变元层，直接回到上一层)(line 1),定义一个空的搜索空间，记录所有未走完所有分支的决定变元(line 2).按顺序将所有未走完所有搜索空间的决定变元记录在nullSearch中(lines 3-5)。如果满足已经搜索到的解超过需要求解的解的数量的一半，且单一分支的决定变元数量达到了一定的数量，随机回溯到一个未走完所有分支的变元层，并记录回溯变元层(lines 6-9)

As shown in Algorithm $\ref{getLevel}$, a heuristic procedure for the random backtracking level of the adaptive DPLL-based algorithm is described. 
First, the backtracking variable level is initialized to 0, i.e., there is no special variable level to go back to, the algorithm goes directly back to the previous level (line \ref{Alg5Line:prelevel}).
Then, an empty search space is defined to record the decision variables that have not gone through all the branches (line \ref{Alg5Line:N}). 
Next, we record decision variables that have not visited all the search space in order in $N$ (lines \ref{Alg5Line:unAssignedVariableBegin}-\ref{Alg5Line:unAssignedVariableEnd}). 
Finally, if it is satisfied that the number of obtained solutions is more than half of the number $maxNum$ of needed solutions, i.e., $2 \times |S| > maxNum$, and the number of decision variables in a single branch reaches a threshold value $Tv$, which $maxNum$ and $Tv$ are empirically set to 30 and 200, then it randomly backtracks to a variable level that has not gone through all branches and record it (lines \ref{Alg5Line:variableLevelBegin}-\ref{Alg5Line:variableLevelEnd}).
%这里的maxNum和Tv的设置是否需要考虑写的更专业点

\begin{algorithm}
    \caption{getLevel$($\sampleSet$,\unAssVarSet,\determin,\prefer)$}
    \label{getLevel}
    \KwIn{Sample set of solution $\sampleSet$, set of unassigned variables $\unAssVarSet$, set of assigned variables $\determin$, assignment weights for each variable $\delta$}
    \KwOut{Select the variable level for backtracking $\prelevel$}
    $\prelevel\gets 0$\label{Alg5Line:prelevel}\\
    $N \gets \emptyset $\label{Alg5Line:N}\\
    \For{$x \in  \determin$}{\label{Alg5Line:unAssignedVariableBegin}
        \If{$x~\text{\rm has not completed all branches}$}{
            $N \gets N \cup\{x\}$\\
        }
    }\label{Alg5Line:unAssignedVariableEnd}
    
    \If{$2\times |\sampleSet|>\maxSolNum$ and $| \determin |>Tv$}{\label{Alg5Line:variableLevelBegin}
        $t \gets $ a randomly picked candidate in $[1,100]$\\
        $\randBackProbability \gets \frac{10\times |\sampleSet|}{\maxSolNum}$\\
        \If{$t<\randBackProbability$}{
            $\prelevel \gets$ a random number in set $N$ \\
        }
    }\label{Alg5Line:variableLevelEnd}
    return $\prelevel$\\
    \BlankLine
\end{algorithm}
    \vspace{-2.7em}

\xsec{Experimental Investigation}\label{sec:experiment}

\zqyAdd{
% In this section, we aim to validate the effectiveness of the proposed algorithm framework SAT-BO. The experiments can be divided into two main parts. In the first part, we mainly verify the effectiveness of the main algorithm modules in SAT-BO, i.e., customized DPLL slover and Bayesian Optimization, in the offline datasets. First we describe the details to construct the offline dataset and conduct detailed ablation study. In the second part, we present the online deployment performance. 

This section is dedicated to evaluating the efficacy of the proposed SAT-BO. Our experimental design comprises two primary parts. The first part focuses on validating the effectiveness of SAT-BO's core algorithmic modules, i.e., the customized DPLL solver and BO, utilizing offline datasets. We describe the details to construct these offline datasets, followed by a comprehensive ablation study to isolate and assess the contribution of each component. Subsequently, the second part presents results in online deployment scenarios. The source code and data have been made available. \footnote{https://github.com/paperpass1024/SAT-BO}}
 %we conduct extensive experiments on 5 real-world business scenarios
%\lmNote{直接介绍算例来自于真实业务场景：part1:业务规则的SAT编码；part2:场景覆盖率源于问题性质的仿真（模拟）}

% \xsubsec{Construction of Vulnerability Coverage Components }\label{sec:simulation }

% In this paper, we employ a polynomial method to artificially construct the function $f(s)=\sum_{i=1}^n g(x_i)$, where $g(x_i)$ stands for the vulnerability coverage of the field where $x_{i}$ becomes the attack field, referred to as the vulnerability coverage component.

% Since the amount of transactions in real-business scenarios is massive, it is impractical to sequentially intercept traffic using attack rules. 
% In this paper, we use simulation to emulate real traffic, thereby setting the vulnerability coverage for each attack rule and evaluating the effectiveness of the proposed algorithm.

\xsubsec{Offline Experiments}\label{sec:offline}
\xsubsubsec{Dataset Construction}
The offline dataset consists of check rules and transaction simulator.
For check rules, we extract 30 verification rules from business cases and transform them to 
SAT instances. The number of variables ranges from 62 to 1341, and the number of clauses ranges from 121 to 2632.

Generally, the transactions of real-business scenarios is characterized by high intensification and relative stability, i.e., most transactions are concentrated on a few variables and the amount of transactions is relatively stable every day.
Based on the characteristics of real business scenarios, and to facilitate calculation, the vulnerability coverage $f(s)$ is decomposed into the sum of the vulnerability coverage $g(x_{i})$ of a series of variables, i.e.,$f(s)=\sum_{i=1}^n g(x_{i})$, $g(x_{i})$ also referred to as the vulnerability coverage component. 
Combining the above characteristics of the real business scenario, this paper adopts the simple but effective polynomial method to simulate the vulnerability coverage component $g(x_{i})$. 
The generic expression for $g(x)$ is shown in the \Equation{} ($\ref{g(x)}$).
\begin{align}\label{g(x)}
%g(x_{i})=\begin{cases}A_{i}&x_{i}=0\\B_{i}&x_{i}=1\end{cases}
g(x_{i})=\begin{cases}1+p_{2i-1}+p_{2i-1}^2&x_{i}=0\\1+p_{2i}+p_{2i}^2&x_{i}=1\end{cases}
\end{align}

For a particular \verificationRule{} encoded into a SAT instance $j$, the best solution found is defined as $\BestSolution{}$.
In order to evaluate the optimality of the solution more clearly, the upper bound on the vulnerability coverage for the SAT instance $j$ is defined to be $C_j=\sum_{i=1}^n \max{\left(g(x_{i})\right)}$.
Thus, our objective is to make the optimal solution $\BestSolution{}$ that is found closer to the upper limit of vulnerability coverage $C_j$.
For the $K$ group \verificationRule{s}, the total vulnerability coverage is $\Theta=\frac{\sum_{j=1}^K \BestSolution{}}{\sum_{j=1}^K C_j}$.

Specifically, we calculate the vulnerability coverage components by means of two representative probabilistic models, including the binomial distribution and power-law distribution models, respectively. 

%假设每个变元都是随机变量，构造一组长度为$2n$的序列$P=\{p_1,p_{2},\cdots,p_{2n-1},p_{2n}\}$，满足概率模型$P\sim B(n,p)$,定义事件$A=p_i\in [0,10]$,事件$B=p_i\in [10,100]$,使得发生事件$A$的概率为$p$,发生事件$B$的概率为$1-p$.
% \par 定义$g(x)$的表达形式如公式$\ref{g(x)Bino}$所示。
\textit{Binomial distribution}:
Assuming that each variable is random, we construct a set of sequences $P=\{p_1,p_{2},\cdots,p_{2n-1},p_{2n}\}$ of length $2n$, which satisfy the binomial distribution model $P\sim B(n,p)$. 
Define the event $A=p_i\in [0,10]$, and the event $B=p_i\in [10,100]$, such that the probability of occurrence of event $A$ is $p$, and the probability of occurrence of event $B$ is $1-p$.
 %The expression of $g(x)$ is shown in the \Equation{} ($\ref{g(x)Bino}$).

%\begin{align}\label{g(x)Bino}
%g(x_{i})=\begin{cases}1+p_{2i-1}+p_{2i-1}^2&x_{i}=0\\1+p_{2i}+p_{2i}^2&x_{i}=1\end{cases}
% \end{align}

% \textit{Half-normal distribution}:
% Similar to the binomial distribution, a set of sequences $Q=\{q_1,q_{2},\cdots,q_{2n-1},q_{2n}\}$ of length $2n$ is constructed randomly, so that $P$ conforms to a normal distribution.
% Specifically, $Q\sim N(\mu,\sigma^2)$, and $\mu$ is the average value of the sequence $Q$.
% Assuming that the probability corresponding to different vulnerability coverage components conforms to a half-normal distribution, i.e., the half-normal distribution probability model needs to satisfy that the vulnerability coverage components of most of the literals are small, and the vulnerability coverage components of a few literals are large.
% We define the sequence $P=\{p_1,\cdots,p_i,\cdots, p_{2n}\}$ of length $2n$ to denote the vulnerability coverage components, and $p_i$ to be presented as in \Equation{} ($\ref{eq:p}$).

% \begin{align}\label{eq:p}
% p_i=\begin{cases}q_i&q_i>\mu \\2\mu-q_{i} &q_i<\mu \end{cases}
%  \end{align}

 %\par 定义$g(x)$的表达形式如公式$\ref{g(x)normal}$所示。
 %The definition $g(x)$ is shown in the \Equation{} ($\ref{eq:g(x)normal}$).
%\begin{align}\label{eq:g(x)normal}
%g(x_{i})=\begin{cases}1+p_{2i-1}+p_{2i-1}^2&x_{i}=0\\1+p_{2i}+p_{2i}^2&x_{i}=1\end{cases} \end{align}

% 本文使用$\mu=0,\sigma\in [300,1000]$的参数范围进行随机构造
% The parameter range for the random construction used in this paper is $\mu=0,\sigma\in [300,1000]$.
\textit{Power-law distribution}:
\yw{
Power-law distribution is a special scenario that may occur in real business scenarios, i.e., the vulnerability coverage component corresponding to most of the variables is 0.
The vulnerability coverage is determined by a small part of the variants, i.e., only a part of the key variables play a role in the decision, and the vulnerability coverage component of such variables is very large.
In order to simulate the business power-law distribution scenario, we define a sequence $P=\{p_1,p_{2},\cdots,p_{2n-1},p_{2n}\}$ of length $2n$, and a subscripted sequence $T=\{t_1,\cdots,t_r\}$ of length $r=\ln{2n}$ that satisfies $\forall t_i\in T, 1\le t_i\le 2n$ and $ 1\le i,j \le r \land i \neq j ,t_i\neq t_j$, such that the sequence $P$ satisfies the \Equation{} ($\ref{eq:highDimen}$).
}
\begin{align}\label{eq:highDimen}
p_i=
\begin{cases}\text{a random number in } [1000,2000]&i\in T\\0 &otherwise \end{cases}
\end{align}
 %\par 定义$g(x)$的表达形式如公式$\ref{eq:g(x)hignDimen}$所示。
% \par The definition $g(x)$ is shown in the \Equation{} ($\ref{eq:g(x)hignDimen}$).
%\begin{align}\label{eq:g(x)hignDimen}
%g(x_{i})=\begin{cases}1+p_{2i-1}+p_{2i-1}^2&x_{i}=0\\1+p_{2i}+p_{2i}^2&x_{i}=1\end{cases}
% \end{align}

\xsubsubsec{Experiment protocol}\label{sec:protocol}
% \par 我们提出的BOBSS算法采用C++和python编程，采用GNU编译器，python版本3.9。所有的实验均在Ubuntu 22.04 LTS(64bit)操作系统下使用Intel(R) Core(TM) i5-8265U CPU。

% \yw{Our proposed \AlgNameAbb{} algorithm is programmed in C++ and compiled with Python v3.9.
% All computation experiments are run on Ubuntu 22.04 LTS (64bit) with Intel (R) Core (TM) i5-8265U CPU.
% }

% In this paper, we conduct experiments on a total of 30 challenging SAT instances transformed by verification rules for 30 complex business cases.
% The number of variables ranges from 62 to 1341, and the number of clauses ranges from 121 to 2632.

Due to the online nature of the problem, the interaction counter between SAT solver and \Baye{} optimization is limited. 
% This paper uses a heuristic method to dynamically allocate the interaction counter to each SAT instance based on the number of variables. 
In the offline experiment, we set the iterative interaction counter as 15.
For 30 SAT instances, one set of simulation scenarios consisting of transactions is randomly generated for each probability model.
The total time required for the SAT solver in the experiment is fixed. If \Baye{} optimization is used, the SAT solver will be given 10 seconds to solve a fixed set of 30 solutions for 15 rounds. The total time limit for the SAT solver is 150 seconds. If \Baye{} is not used, the SAT solver will be given 150 seconds to solve 450 solutions separately before calculating the total coverage. 
% In fact, our verification can be completed in one day, so this experiment is not sensitive to time overhead. Our current time setting is fully in line with business requirements.
We performed three independent runs of each SAT instance in each scenario and calculated the vulnerability coverage percentages $\Theta$ for each instance.

\xsubsubsec{Results and Ablation Study}\label{sec:results}

In this subsection, we comprehensively study the performance of \AlgNameAbb{} and report the detailed results of all SAT instances on two sets of scenarios generated by two constructed vulnerability coverage component functions which are binomial and power-law. 
\begin{table}[ht]
\vspace{-0.8em}
\centering
\setlength{\tabcolsep}{2.3pt}
    \footnotesize
    %\scriptsize
    \begin{tabular}{*{11}{c}}
    \toprule
     \multirow{2}*{Algorithm}&\multirow{2}*{RB}&\multirow{2}*{RFDV}&\multirow{2}*{RFIV}&\multirow{2}*{BO}&\multicolumn{2}{c}{binomial}&\multicolumn{2}{c}{power-law}\\
    \cmidrule(lr){6-7}		\cmidrule(lr){8-9}		\cmidrule(lr){10-11}
    ~&~&~&~&
    ~& $f_{best}$&$f_{avg}$& $f_{best}$&$f_{avg}$\\
    \midrule
    Random&~&~&~&~&21  &19.56  & 18.58 & 17.25 \\
    \AlgNameAbb{0}&~&~ &~&\checkmark &86.68&82.11&99.60&98.46\\
    \AlgNameAbb{1}&~&\checkmark &\checkmark&\checkmark&88.90&83.71&99.13&98.62\\
    \AlgNameAbb{2}&\checkmark&~&\checkmark&\checkmark&90.16&85.56&\textbf{99.61}&98.94\\
    \AlgNameAbb{3}&\checkmark &\checkmark &~&\checkmark&90.30&84.74&98.58&98.44\\
    \AlgNameAbb{4}&\checkmark &\checkmark &\checkmark&~&71.67&58.64&69.51&55.87\\
    WalkSAT-BO&~&~ &~&\checkmark &83.51&80.08&88.58&81.78\\
    \AlgNameAbb{}&\checkmark &\checkmark &\checkmark&\checkmark&\textbf{92.75}&\textbf{87.40}&99.60&\textbf{99.28}\\

    \bottomrule
    \end{tabular}
    \caption{Overall computational results on all test scenarios.}
    \label{tab:Total}

\end{table}
\vspace{-1.0em}
 
To confirm the effectiveness of our proposed three \filpStrageName, including \randBack{}, \randFilpCandid{} and \randFilpIndepend{}, we also evaluate \AlgNameAbb{} with five alternative versions named \AlgNameAbb{0}, \AlgNameAbb{1}, \AlgNameAbb{2}, \AlgNameAbb{3} and \AlgNameAbb{4} respectively. Meanwhile, to confirm the effectiveness of DPLL SAT algorithm, we use WalkSAT as the representative of local search SAT algorithm to make a comparison. Here, we name the alternative version solver using WalkSAT as WalkSAT-BO.
To substantiate the efficacy of our proposed approach, we employ a random attack method, denoted as the Random (solver), to serve as a baseline for comparative performance analysis against SAT-BO. 
% It is important to note that this random solver represents a straightforward  vulnerability detection method that has been widely utilized by researchers in the field.

As shown in \Tab{} \ref{tab:Total}, these variant solvers have the same components except for the configuration of the \filpStrageName{}, \Baye{} optimization and the framework of SAT solver.
\begin{itemize}
    \item \textit{Random}: Randomly determine the value of each variable to see if it satisfies the solution, and continuously find a sufficient number of SAT solutions.
    \item \textit{\AlgNameAbb{0}}: Disable all \filpStrageName.
    \item \textit{\AlgNameAbb{1}}: Disable the \randBack{}.
    \item \textit{\AlgNameAbb{2}}: Disable the \randFilpCandid{}.
    \item \textit{\AlgNameAbb{3}}: Disable the \randFilpIndepend{}.
    \item \textit{\AlgNameAbb{4}}: Disable the \Baye{} optimization.
    \item \textit{WalkSAT-BO}: 
    Replaced the DPLL SAT algorithm based on tree search with the WalkSAT algorithm based on a local search framework. 
    % Here, Walksat uses the assignment tendency provided by \Baye{} optimization as the initial value to complete the interaction with \Baye{} optimization.
\end{itemize}

We provide evaluations of the performance of seven versions on all SAT instances and report the detailed results in \Tab{s} \ref{tab:Total}, \ref{tab:Binomial} and \ref{tab:Power-Law}.
Column ``Algorithm'' represents the names of versions of algorithms. 
Column ``Instance'' gives the names of the SAT instances.
Columns ``$f_{best}$'' and ``$f_{avg}$'' represent the best and average vulnerability coverage percentage obtained by these algorithms in all independent runs.
Row “\#best” indicates the number of instances for which each algorithm obtains the best results among all algorithms.
% Furthermore, \Fig{} \ref{fig:analysis} depicts the vulnerability coverage gaps between the results obtained by the corresponding algorithms and the best results on all algorithms.
% Each point $(x, y)$ on the curves denotes that the gap between the vulnerability coverage of the current solution and the best known one of ``$f_{best}$'' or ``$f_{avg}$'' is $y$ at $x$ instance.

\Tab{} \ref{tab:Total} shows the overall results on two sets of scenarios obtained by \AlgNameAbb{}, \AlgNameAbb{0}, \AlgNameAbb{1}, \AlgNameAbb{2}, \AlgNameAbb{3}, \AlgNameAbb{4} , Random and WalkSAT-BO.
We can observe that \AlgNameAbb {} is able to obtain the optimal vulnerability coverage in all sets of scenarios except for the scenario named power-law. But in the power-law scenario, \AlgNameAbb{} is only 0.01 below the optimal vulnerability coverage.

\Tab{s} \ref{tab:Binomial} to \ref{tab:Power-Law} compare the experimental results of \AlgNameAbb{} with its variants.
From \Tab{s} \ref{tab:Binomial} to \ref{tab:Power-Law}, one can observe that \AlgNameAbb{} is the best-performing algorithm among the seven versions in all instance of two scenarios.
In terms of the average results, for the two different scenarios, \AlgNameAbb{} obtains better results on all 30 SAT instances.
In terms of the best results, \AlgNameAbb{} obtains the best results for 13 (out of 30) SAT instances for binomial distribution scenario and obtains the best results for 26 (out of 30) SAT instances for power-law distribution scenario, while other variants obtain the best results on only no more than 12 instances for binomial distribution scenario.
% In addition, from \Fig{} \ref{fig:analysis}, we can observe that as for the results of versions which disable some strategies, \AlgNameAbb{} keeps an advantage of about 10\% in the best and average results.
% Even on some instances, \AlgNameAbb{} has an advantage of over 35\%, such as v79\_c334 of \AlgNameAbb{4} in \Fig{} \ref{fig:analysis} (e).

%\AlgNameAbb{} obtains the best results for 19 (out of 30) SAT instances for half-normal distribution scenario

For the solving efficiency, Tables \ref{tab:Random} shows the number of satisfied solutions obtained by the random solver in 150 seconds, while the DPLL SAT solver can obtain at least 30 solutions in 1 second. It can be seen that the solving efficiency of random is much lower than that of SAT solver. 

% Tables \ref{tab:all-random} shows the comparison of experimental results between random solver and SAT-BO. It can be seen that with the increase of the difficulty of the instance, the result of random solver decreases significantly, and the stability is far lower than that of SAT-BO.

% These results confirm that the three \filpStrageName{} and \Baye{} optimization are essential for \AlgNameAbb{} to obtain competitive results. In sum, these statistics demonstrate that our \AlgNameAbb{} is highly effective, efficient, and robust for solving the \proName.

\xsubsec{Online Experiments}\label{sec:results}
% setting/metric/results
% To evaluate the actual online performance of SAT-BO， we conducted a 15-day online test in on business scenario. In the online production environment, we focused on the attack success ratio, defined as the proportion of synthetic transactions that either not triggered or passed the verification rules relative to the total number of synthetic transactions. The results demonstrate that SAT-BO yields a significant improvement of \textbf{38.18\%} in the attack success ratio compared to the baseline in the online production environment. Note that the baseline generates synthetic transactions by randomly selecting values within the range of the given tampered field with the intention of producing transactions that constitute vulnerabilities. 
% Moreover, SAT-BO could generate new verification rules to avoid vulnerabilities. But baseline cannot do that.

To assess the actual online performance of SAT-BO, we conducted 15-day online evaluation within one business scenario in the cooperationg company. In the production environment, our primary focus was on the attack success ratio, defined as the proportion of synthetic transactions that either not triggered or passed the verification rules, relative to the total number of synthetic transactions generated. The results demonstrated that SAT-BO achieved a statistically improvement of \textbf{38.18\%} in the attack success ratio compared to the baseline method in the online production environment. It is important to note that the baseline approach generates synthetic transactions by randomly selecting values within the range of the given tampered fields, with the intention of producing transactions that exploit potential vulnerabilities.

Furthermore, the attack rules generated by SAT-BO with vulnerability coverage percentage larger than 30\% are converted to verification rules, which is a
feature notably absent in the baseline method. This additional functionality underscores the adaptive and proactive nature of SAT-BO in enhancing system security, offering a distinct advantage over traditional approaches.

Regarding computational efficiency, SAT-BO gives solutions for individual SAT instances within seconds. The total processing time for the entire set of SAT instances remains under one hour when utilizing 32 CPU cores and 128 GB of memory.

\xsec{Conclusion}

% \zqyAdd{In this paper, we propose \AlgNameAbb{} method providing theoretical foundation for generating vulnerability-free verification rules, which integrates customized adaptive DPLL SAT solver and BO to find the attack rule with highest vulnerability coverage.
% We perform comprehensive experiments to show the impact of the three \filpStrageName{} in SAT solver and BO module of \AlgNameAbb. The experiments show that the three \filpStrageName{} and BO significant roles in reaching a trade-off between exploration and exploitation of the search, thus making \AlgNameAbb{} powerful and robust.
% Furthermore, SAT-BO has been deployed on the financial security platform for Alipay in AntGroup. }

In this paper, we propose the \AlgNameAbb{} method, which provides the robust theoretical foundation for generating vulnerability-free verification rules. This approach integrates the customized adaptive DPLL SAT solver with BO to identify attack rules with maximal vulnerability coverage. We conduct comprehensive experiments to evaluate the impact of the three \filpStrageName{} within the SAT solver and the BO module of \AlgNameAbb{}. The results demonstrate that these \filpStrageName{}, in conjunction with BO, play crucial roles in achieving optimal balance between exploration and exploitation during the search process, thereby enhancing the efficacy and robustness of \AlgNameAbb{}. Moreover, SAT-BO has been deployed on the financial security platform in the cooperating company, enhancing the verification rules that safeguard transactions.